\documentclass{emulateapj}

\usepackage{natbib}
\usepackage{subfigure}
\usepackage{lscape}

\newcommand{\captionfonts}{\small}

\makeatletter  
\long\def\@makecaption#1#2{%
  \vskip\abovecaptionskip
  \sbox\@tempboxa{{\captionfonts #1: #2}}%
  \ifdim \wd\@tempboxa >\hsize
    {\captionfonts #1: #2\par}
  \else
    \hbox to\hsize{\hfil\box\@tempboxa\hfil}%
    \fi
  \vskip\belowcaptionskip}
\makeatother   

\shorttitle{Blue-Sequence E/S0s}
\shortauthors{Wei et al.}

\defcitealias{kannappan09a}{KGB}
\defcitealias{kannappan09b}{K09b}

\begin{document}

\title{Gas Mass Fractions and Star Formation in Blue-Sequence E/S0 Galaxies}

\author{Lisa H. Wei\altaffilmark{1}, Sheila
  J. Kannappan\altaffilmark{2}, Stuart N. Vogel\altaffilmark{1},
  Andrew J. Baker\altaffilmark{3}}
\altaffiltext{1}{Department of Astronomy, University of Maryland, College Park, MD 20742-2421}
\altaffiltext{2}{Department of Physics and Astronomy, University of North
  Carolina, 290 Phillips Hall CB 3255, Chapel Hill, NC 27599-3255}
\altaffiltext{3}{Department of Physics and Astronomy, Rutgers, the State
  University of New Jersey, 136 Frelinghuysen Road, Piscataway, NJ 08854-8019}

\begin{abstract}
  Recent work has identified a population of low-redshift E/S0
  galaxies that lie on the blue sequence in color vs. stellar mass
  parameter space, where spiral galaxies typically reside. While
  high-mass blue-sequence E/S0s often resemble young merger or
  interaction remnants likely to fade to the red sequence, we focus on
  blue-sequence E/S0s with lower stellar masses (${M_*} <$ a few
  $\times 10^{10}\,M_{\odot}$), which are characterized by fairly
  regular morphologies and low-density field environments where fresh
  gas infall is possible. This population may provide an evolutionary
  link between early-type galaxies and spirals through disk
  regrowth. Focusing on atomic gas reservoirs, we present new GBT HI
  data for 27 E/S0s on both sequences as well as a complete tabulation
  of archival HI data for other galaxies in the Nearby Field Galaxy
  Survey. Normalized to stellar mass, the atomic gas masses for 12 of
  the 14 blue-sequence E/S0s range from 0.1 to $>$1.0, demonstrating
  that morphological transformation is possible if the detected gas
  can be converted into stars. These gas-to-stellar mass ratios are
  comparable to those of spiral and irregular galaxies and have a
  similar dependence on stellar mass. Assuming that the HI is
  accessible for star formation, we find that many of our
  blue-sequence E/S0s can increase in stellar mass by 10--60\% in 3
  Gyr in both of two limiting scenarios, exponentially declining star
  formation (i.e., closed box) and constant star formation (i.e.,
  allowing gas infall). In a constant star formation scenario, about
  half of the blue-sequence E/S0s require fresh gas infall on a
  timescale of $\lesssim$3 Gyr to avoid exhausting their atomic gas
  reservoirs and evolving to the red sequence. We present evidence
  that star formation in these galaxies is bursty and likely involves
  externally triggered gas inflows. Our analysis suggests that most
  blue-sequence E/S0s are indeed capable of substantial stellar disk
  growth on relatively short timescales.
\end{abstract}


\section{Introduction} 
Current models of galaxy formation and evolution favor hierarchical
growth of galaxies from smaller systems (e.g.,
\citealt{white91,somerville99,bower06}). Within the paradigm of
hierarchical galaxy formation, galaxies evolve along the Hubble
sequence, transforming back and forth between E/S0 and
spiral/irregular morphology through a series of quiescent and violent
periods.

Recognition that galaxies can transform from late to early type dates
back at least to \citet{toomre72}, who proposed that elliptical
galaxies can form from mergers of similar mass late-type
galaxies. Recent simulations find that while similar mass (1:1--3:1)
mergers of disk galaxies result in classical ellipticals, unequal mass
(4.5:1--10:1) or gas-rich mergers of galaxies result in S0s and hybrid
galaxies with properties of both early and late types
\citep{bekki98,naab06,bournaud05}. Substantial observational evidence
confirms that early-type galaxies can form through major mergers
(e.g., \citealt{schweizer92,vandokkum05}). In agreement with the
simulations, \citet{emsellem07} argue that larger, slow-rotating
elliptical galaxies form from dry major mergers while fainter,
fast-rotating ellipticals and S0s seem to form from minor mergers or
very gas-rich major mergers.

Can galaxies evolve in the other direction, from early to late type?
Simulations suggest that galaxy evolution in this direction involves
the regrowth of stellar disks
\citep{steinmetz02,governato07}. Morphological transformation from
early to late type requires that several conditions be met, which
recent studies have found satisfied in some early-type galaxies.

First, the transformation from early- to late-type requires a
substantial reservoir of cold gas --- the raw material for disk
regrowth. Historically, early-type galaxies were thought to be
gas-poor \citep{faber76,knapp78}. Subsequent surveys, however, found
that the ratio of gas mass to blue luminosity (${M_{\rm HI}/L_{B}}$)
in early-type galaxies ranges from upper limits of ${M_{\rm HI}/L_{B}}
\sim 0.009\, {M_{\odot}/L_{\odot}}$ to measured ${M_{\rm HI}/L_{B}}
\sim 2.7\, {M_{\odot}/L_{\odot}}$ for large atomic gas reservoirs
similar to those of spiral galaxies
\citep{hawarden81,knapp85,wardle86,sadler00,oosterloo02}. The 70\% HI
detection rate by \citet{morganti06} suggests that atomic gas
reservoirs are actually relatively common in field early-type
galaxies. Recent surveys also find a significant amount of molecular
gas ($10^{7}$--$10^{9} M_{\odot}$) in 28--78\% of early-type galaxies,
depending on the survey
\citep{lees91,knapp96,welch03,sage07,combes07}. The large range in
cold gas content hints that there may be distinct sub-populations of
early-type galaxies --- the conventional red and dead early types, and
a population that is still accreting gas and forming stars.

In addition to the {\it presence} of cold gas, the distribution of the
gas is important in assessing the potential for disk regrowth. The
existence of giant HI disks and rings \citep{morganti97,serra07}, as
predicted by simulations of mergers between gas-rich galaxies
\citep{barnes02}, makes it possible for stellar disks to form from
already-present gas disks. The HI structures around some early-types
galaxies are known to have regular velocity fields, with a continuity
between ionized and neutral gas
\citep{vangorkom86,schweizer89,schiminovich95,morganti06}. \citet{hibbard96}
find a rotationally supported HI disk in the elliptical galaxy
NGC~520, which, if star formation is triggered in the disk, may
generate a late-type galaxy. Kiloparsec-scale disks of regularly
rotating molecular gas have also been found in the center of some
early-type galaxies \citep{young02,young05,young08,crocker08}. All
these observations suggest that the gas disks around some, possibly
most early-type galaxies have reached an equilibrium arrangement
suitable for disk growth.

While the presence of cold gas disks in an equilibrium configuration
is necessary for stellar disk growth, it does not imply that star
formation is actually occurring.  A key question is whether E/S0
galaxies with such disks are currently evolving morphologically, and
whether such galaxies constitute a significant fraction of the
early-type galaxy population. To answer these questions, one must be
able to distinguish between old, gas-poor early-type galaxies and
those undergoing morphological transformation via disk regrowth.

\citet*[hereafter KGB]{kannappan09a} recently identified a population
of E/S0s that reside alongside spiral galaxies on the blue sequence in
color vs. stellar mass parameter space. They argue that among these
``blue-sequence E/S0s,'' a subset with low-to-intermediate masses may
represent the missing link --- i.e., galaxies that are actively
(re)growing stellar disks and plausibly transitioning from early to
late type. \citetalias{kannappan09a} find that these blue-sequence
E/S0s fall between spirals and red-sequence E/S0s in scaling relations
(stellar mass $M_*$ vs.\ radius and vs.\ velocity dispersion
$\sigma$), implying that blue-sequence E/S0s might form a transitional
population between the two groups. Compared to conventional
red-sequence E/S0s, blue-sequence E/S0s consistently have bluer outer
disks, and often bluer centers as well --- suggesting on-going star
formation in both disks and disky bulges. Further supporting this
disk-building picture, \citetalias{kannappan09a} find evidence for
kinematically distinct disks (i.e., counterrotating or polar) in a
notable sub-population of blue-sequence E/S0s.

Unlike red-sequence E/S0s, whose mass function peaks at high masses,
blue-sequence E/S0s are rare for $M_* >$ 1--2$\,\times
10^{11}\,M_{\odot}$, and are common only for ${M_* < 3 \times
  10^{10}\,M_{\odot}}$ \citepalias{kannappan09a}. Their abundance
increase sharply to 20--30\% below $M_* \lesssim 5 \times
10^9\,M_{\odot}$, coincident with the mass threshold below which
galaxies become notably more gas rich (\citealt{kannappan08conf},
based on \citealt{kannappan04}). At intermediate masses between
$10^{10}$ and $10^{11}\,M_{\odot}$, \citetalias{kannappan09a} find
that blue-sequence E/S0s include examples of both major mergers that
are likely to fade onto the red sequence after exhausting their gas,
and also settled galaxies that may be evolving toward later-type
morphologies. The former dominate at higher masses, and the latter at
lower masses.

One outstanding question that remains is the extent of disk growth
possible in low-to-intermediate mass blue-sequence E/S0s. In this
paper, we approach this question from the point of view of atomic gas
reservoirs. \citetalias{kannappan09a} addressed this question with
limited archival data for the Nearby Field Galaxy Survey (NFGS,
\citealt{jansen00a}), finding preliminary evidence for gas reservoirs
comparable to spiral galaxies. In this paper we present more complete,
higher-quality data for all NFGS E/S0s with $M_* <
4\times10^{10}\,M_{\odot}$ and a sampling of more massive E/S0s. We
compare the atomic gas masses (normalized to ${M_*}$) of blue-sequence
E/S0s with those of other galaxies, and we examine the possible
fractional growth in ${M_*}$ given current rates of star
formation. The question of how efficiently the atomic gas might flow
inward and condense into molecular gas is beyond the scope of this
paper; however, we discuss preliminary evidence from our own and
others' work that suggests efficient conversion is plausible
(elaborated in a forthcoming paper: \citealt[hereafter
K09b]{kannappan09b}).

Section \ref{section.sample} describes our statistically
representative sample of red- and blue-sequence E/S0s from the NFGS,
and presents new Green Bank Telescope\footnote{The National Radio
  Astronomy Observatory is a facility of the National Science
  Foundation operated under cooperative agreement by Associated
  Universities, Inc.} (GBT) data for the galaxies with $M_* <
4\times10^{10}\,M_{\odot}$, which are expected to show the most disk
growth \citepalias{kannappan09a}. We also present a tabulation of HI
data for the full NFGS. Section \ref{section.gasreservoirs} compares
atomic gas masses of blue-sequence E/S0s with those of red-sequence
E/S0s and galaxies of later-type morphology within the NFGS. We also
compare to the sample of \citet{sage06}. Section
\ref{section.starformation} examines the fractional stellar mass
growth possible for blue-sequence E/S0s given the current star
formation rate, in two limiting scenarios --- constant (i.e., allowing
gas infall) and exponentially declining (i.e., closed box) star
formation. In Section \ref{section.sfmechanism}, we discuss gas
exhaustion and gas inflow timescales, and we examine evidence for
bursty star formation in blue-sequence E/S0s, which likely implies
efficient conversion of HI to ${\rm H_2}$. We conclude with a
discussion of the evolutionary fates of blue-sequence E/S0s.

Appendix \ref{section.appendix} details features of our new HI data
for interesting individual galaxies. In this paper, we assume $H_0$ =
70 ${\rm km\, s^{-1}\, Mpc^{-1}}$.

\section{Sample and Data}\label{section.sample}
\subsection{Sample}
\subsubsection{NFGS Sample and Data}
To understand the role of blue-sequence E/S0s in the morphological
evolution of galaxies, we need to examine these galaxies alongside
different types of galaxies in various stages of evolution. The NFGS
provides an ideal parent sample for such study, spanning the natural
diversity of galaxies in the local universe in terms of mass,
luminosity, and morphological type. \citet{jansen00a} selected the
NFGS galaxies from the CfA redshift catalog, which they binned by
absolute magnitude and sub-binned by morphological type. After
applying a luminosity-dependent minimum redshift to avoid galaxies of
a large angular size, \citet{jansen00a} chose every Nth galaxy from
each bin, scaling N to approximate the local galaxy luminosity
function. The resulting NFGS spans the full range of morphological
types and eight magnitudes in luminosity, providing a distribution of
galaxies that is statistically consistent with that of the local
universe.

Archival NFGS data include $UBR$ photometry \citep{jansen00b},
integrated spectrophotometry \citep{jansen00a}, and ionized gas and
stellar kinematic data \citep{kannappan01,kannappan02}. All NFGS
galaxies also have $JHK$ photometry from 2MASS \citep{skrutskie06}.

Stellar masses are estimated by fitting stellar population models to
$UBRJHK$ photometry and integrated spectrophotometry as described in
\citetalias{kannappan09a} (updating \citealt{kannappan07}; see also
\citealt{kannappan08conf}). This limits the ``full NFGS sample'' to
176 galaxies for which stellar masses are available. For consistency
with the red/blue sequence dividing line of \citetalias{kannappan09a},
we use $U-R$ colors with those authors' extinction corrections and
$k$-corrections. Total magnitudes are also extinction corrected. We
also use star formation rates calculated by \citetalias{kannappan09a}
from extinction-corrected H$\alpha$ spectral line data, integrated by
scanning the slit across each galaxy and calibrated against IRAS-based
star formation rates (SFRs) following \citet{kewley02}\footnote{The
  \citet{kewley02} SFRs were scaled to the same IMF used in the
  stellar mass estimation by \citetalias{kannappan09a}.}.

\subsubsection{Sub-${M_b}$ E/S0 Sample}\label{section.focus}
\begin{figure}[!h]
\includegraphics[scale=0.7,angle=0]{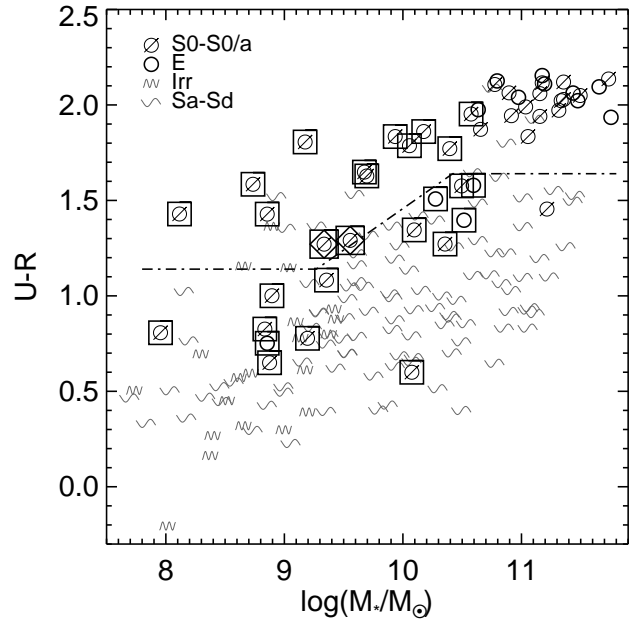}
\caption{$U-R$ color vs. stellar mass for galaxies in the Nearby Field
  Galaxies Survey (NFGS, \citealt{jansen00a}). Symbols denote
  morphological type classified by eye using monochrome $B$ or $g$
  band images \citepalias{kannappan09a}. The red sequence, the main
  locus of high-mass and/or cluster E/S0s, lies above the dashed line,
  while the blue-sequence, typically populated by spiral galaxies,
  lies below. Note the presence of a population of galaxies with
  early-type morphology in the region of color vs. stellar mass space
  populated by spirals. For consistency with the red/blue sequence
  dividing line of \citetalias{kannappan09a}, we use $U-R$ colors with
  those authors' extinction corrections and k-corrections. The
  sub-${M_b}$ E/S0 sample of galaxies are boxed, with the two
  mid-sequence E/S0s on the borderline between the red and blue
  sequences also enclosed in diamonds. Note that in this paper
  ``sub-$M_b$'' means $M_* < 4 \times 10^{10}\,M_{\odot}$, i.e. we use
  a cutoff slightly above $ M_b = 3 \times 10^{10}\,M_{\odot}$, due to
  our original sample selection.\label{fig.nfgssample}}
\end{figure}

We consider a subsample of NFGS E/S0 galaxies for detailed study of
the ${M_*}$ regime where \citetalias{kannappan09a} report abundant
blue-sequence E/S0s. Our focus sample includes all 14 blue-sequence
E/S0s with ${M_* \le 4 \times 10^{10}\,M_{\odot}}$, with the limit
chosen where the blue-sequence E/S0s tail off (Figure
\ref{fig.nfgssample}). To make a fair comparison, we include all 11
NFGS red-sequence E/S0s with ${M_* \le 4 \times
  10^{10}\,M_{\odot}}$. We also include two galaxies that lie on the
dividing line between the red and blue sequences, which we will refer
to as ``mid-sequence'' E/S0s following the naming convention of
\citetalias{kannappan09a}. Our cutoff mass is very close to the
bimodality mass (${M_b \sim 3 \times 10^{10}\,M_{\odot}}$) discussed
in \citetalias{kannappan09a}, so henceforth we refer to this sample as
the ``sub-${M_b}$'' E/S0s.

The sample mass limit $M_* \lesssim M_b$ excludes a large population
of high-mass red-sequence E/S0s; this is appropriate because many
properties scale with stellar mass, and including the high-mass
galaxies would bias statistical comparison. Figure
\ref{fig.nfgssample} shows the sub-${M_b}$ sample in $U-R$ color
vs. stellar mass parameter space, illustrating similar mass
distributions with good coverage of both the red and blue
sequences. \citetalias{kannappan09a} divide the two sequences with the
dashed line in Figure \ref{fig.nfgssample}, which is chosen with
respect to the locus that hugs the upper boundary of the distribution
of most late-type galaxies. The line levels out at $U-R$ values of
1.14 and 1.64, in agreement with \citet{baldry04}. The rest of the
NFGS (spirals and irregulars) is plotted in the background in the same
figure for comparison. Because the full NFGS was selected to be
broadly representative of the local universe, we expect our
sub-${M_b}$ E/S0 sample to encompass a wide range of evolutionary
stages as well.

\subsection{HI Data}
We present the compilation of HI data for the full NFGS in Table
\ref{table.nfgshi}.

\subsubsection{HyperLeda and Literature}
HI fluxes are available for most of the NFGS from the HyperLeda
database \citep{paturel03b}, which consists of data compiled from the
literature. Because the HyperLeda database is compiled from
observations made at different telescopes by different observers,
\citeauthor{paturel03b} homogenize the HI data to account for
differences in observational parameters such as beamsize, spectral
resolution, and flux scale.

For NFGS galaxies lacking HI data in HyperLeda, we gather data from
the literature when possible. We obtain upper limits for three
galaxies from \citet{huchtmeier89}, four HI fluxes and one upper limit
from the Cornell EGG HI Digital Archive \citep{springob05}, and six
upper limits and one flux measurement from the HI Parkes All Sky
Survey (HIPASS, \citealt{hipass01}). Because these measurements are
often upper limits or taken at telescopes not included in the
homogenization effort of \citet{paturel03b}, the fluxes for these
galaxies are not homogenized to the HyperLeda dataset.

\subsubsection{New GBT HI Observations}
Good quality literature data are lacking for many early-type galaxies
in the NFGS, so we have obtained GBT HI observations for many of these
galaxies, with priority for our sample of sub-${M_b}$ E/S0s. The
observations were obtained with the GBT Spectrometer in L-band, with
50 MHz bandwidth, one spectral window, and nine sampling levels, in
ten minute on-off source pairs (five minutes per position) during
March and October of 2007. The total on-source time for each galaxy
was determined during the observing runs based on the strength of the
HI emission relative to the noise.

The spectra were reduced using GBTIDL \citep{gbtidl}. Individual 30
second records with large harmonic radio frequency interference (RFI)
were flagged, and persistent RFI spikes near the velocity range of the
galaxy were interpolated across in a few cases. The scans were
accumulated and averaged for all data for an individual galaxy, and a
polynomial of order $\leq$5 was fitted over a range of $\sim$20 MHz
to subtract the baseline. Hanning and fourth order boxcar smoothing,
with decimation, were then applied to all the baseline-subtracted
data, resulting in channel resolution of 0.0244 MHz ($\sim$5 ${\rm
km\,s^{-1}}$). Flux calibration is derived from simultaneous
observations of an internal noise diode whose intensity is stable.

Figures \ref{fig.gbtspectra_red} and \ref{fig.gbtspectra_blue} present
the spectra of red-, blue-, and mid-sequence E/S0s obtained with the
GBT, and Table \ref{table.newhi} lists the observational parameters
and measured quantities. ~Col.~(6) lists the total on-source time in
seconds; ~Col.~(7) is the heliocentric recession velocity measured at
the mid-point of 20\% flux after excluding companions ($V_{\odot}$,
see \S \ref{section.companions}). ~Cols.~(8) and (9) are the HI
line widths measured at the 20\% and 50\% level ($W_{20},
W_{50}$). ~Col.~(10) gives the velocity-integrated flux and error
($f_{\rm HI}, \sigma_{f_{\rm HI}}$), both in Jy ${\rm km\,s^{-1}}$, and
~Col.~(11) gives the dispersion of the baseline channels measured in a
line-free part of the spectrum ($\sigma_{\rm chan}$) in mJy.
 
We estimate the error of our flux measurements following
\citet{schneider86,schneider90}, who derived the following analytical
expression for uncertainty in total HI flux ($\sigma_{f_{\rm HI}}$):

\begin{equation}
  \sigma_{f_{\rm HI}} = 2\sigma_{\rm chan}\sqrt{1.2 W_{20} \Delta V} {\,\,\,\rm Jy\, km\,s^{-1}}
\end{equation}

\noindent where $\sigma_{\rm chan}$ is the rms dispersion of the baseline
in Jy and $\Delta V$ is the velocity resolution of the spectrum in
${\rm km\,s^{-1}}$. The errors in the measured heliocentric velocity
($\sigma_{\rm V_{\odot}}$) and velocity widths ($\sigma_{W_{50}},
\sigma_{W_{20}}$) are estimated following \citet{fouque90}, using:

\begin{equation}
\sigma_{V_{\odot}} = 4\sqrt{\Delta V\, P}\,\,({\rm
  S/N})^{-1}\,\, {\rm \, km\,s^{-1}}
\end{equation}

\begin{equation}
\sigma_{W_{50}} = 2 \sigma_{V_{\odot}}\,\,{\rm \, km\,s^{-1}}
\end{equation}

\begin{equation}
\sigma_{W_{20}} = 3 \sigma_{V_{\odot}}\,\,{\rm \, km\,s^{-1}}
\end{equation}

\noindent where $P$ is the steepness of the profile,
$(W_{20}-W_{50})/2$, and S/N is the ratio of the peak signal to
$\sigma_{\rm chan}$.

We cannot homogenize our HI fluxes to the HyperLeda system of
\citet{paturel03b}, because these authors do not include the GBT in
their homogenization calculations. However, because of the large beam
of the GBT at 21cm ($\sim$9$\arcmin$) compared to the small optical
sizes of our galaxies ($\sim$2$\arcmin$), no beam-filling correction
is needed. Also, flux calibration of GBT HI spectral line data is
extremely stable and accurate, so large offsets in the flux scale
between our GBT data and the HyperLeda data are unlikely. In fact,
comparison between our new GBT HI data and existing HI data from
HyperLeda for 11 galaxies show that over half have GBT fluxes within
20\% of published HI fluxes, even though the HyperLeda data are of
poorer quality. In all 11 cases the S/N of the GBT HI data is better
than that of the published data. We discuss the HI fluxes and line
profiles of individual galaxies in further detail in Appendix A.

\subsubsection{Companions}\label{section.companions}
\begin{figure*}
\hspace{1cm}
\includegraphics[scale=.8]{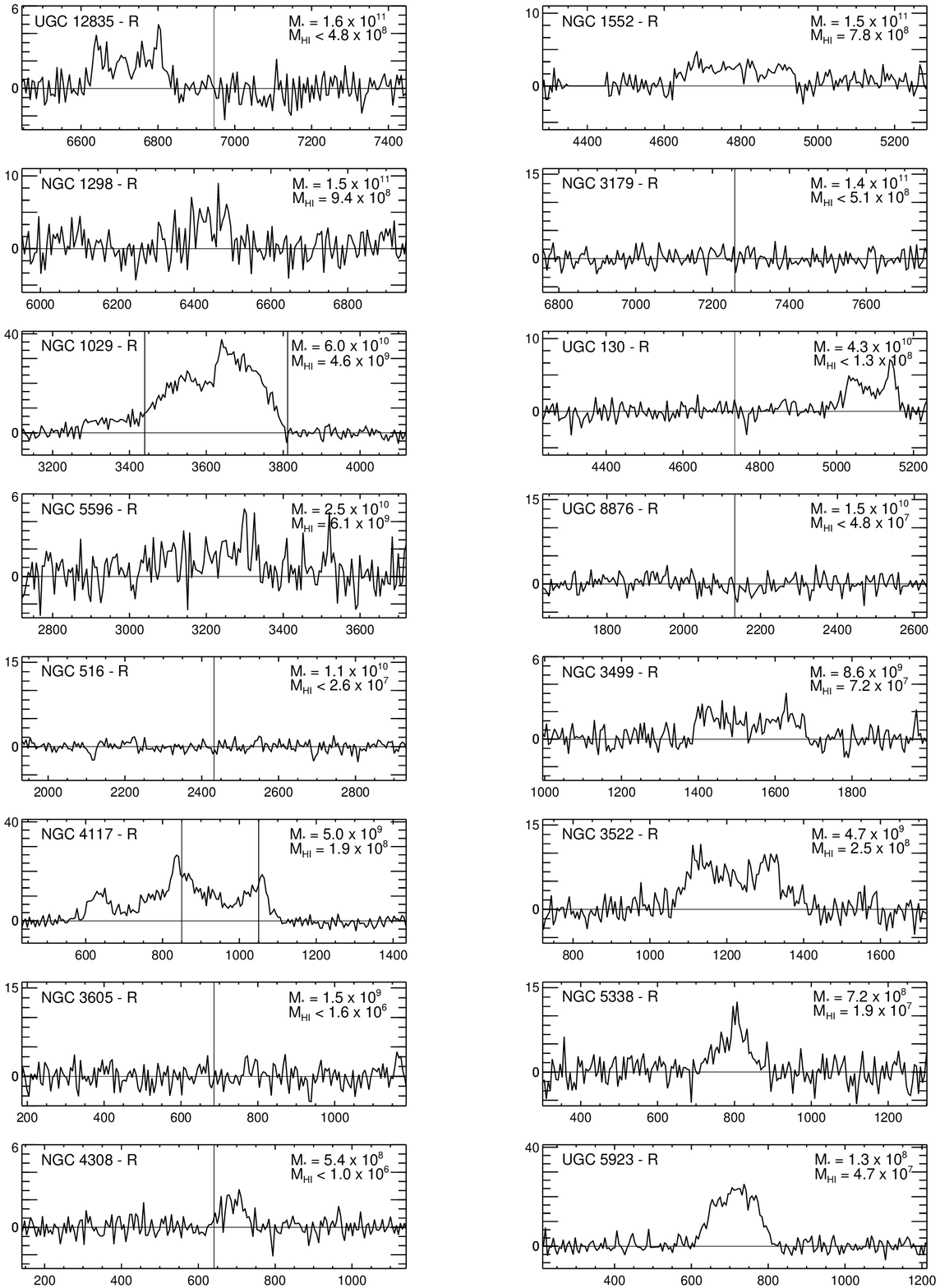}
\vspace{1cm}
\caption{HI spectra of red-sequence E/S0s from the NFGS observed with
  the GBT. The galaxies are ordered left to right, top to bottom by
  decreasing stellar mass. Double vertical lines indicate the ranges
  of ionized-gas rotation (or stellar rotation, if ionized gas data
  are not available) between which we measure the HI flux to exclude
  companion flux. Please see Appendix A for more details on individual
  galaxies. Single vertical lines mark the optical velocities for the
  target galaxies with HI upper limits. NGC~4308 is considered a
  non-detection because the measured HI velocity coincides exactly
  with the optical velocity of a nearby companion. Assigning the gas
  to NGC~4308 would not change any results since the measured HI flux
  would still be extremely low for this galaxy's stellar
  mass.\label{fig.gbtspectra_red}}
\end{figure*}

\begin{figure*}
\hspace{1cm}
\includegraphics[scale=.8]{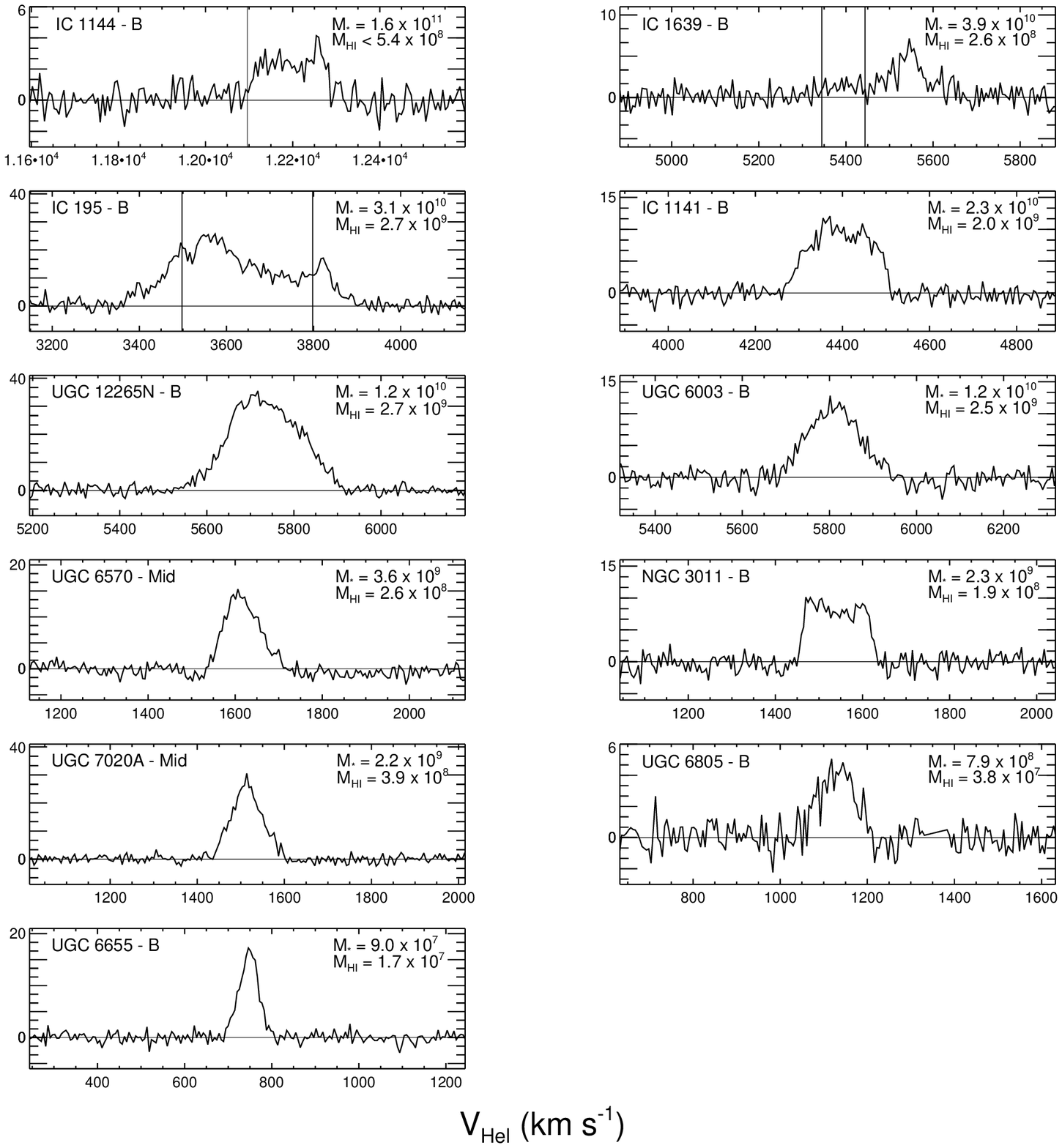}
\caption{HI spectra of blue- and mid-sequence E/S0s from the NFGS
  observed with the GBT. The galaxies are ordered left to right, top
  to bottom by decreasing stellar mass. Double vertical lines indicate
  the ranges of ionized-gas rotation (or stellar rotation, if ionized
  gas data are not available) between which we measure the HI flux to
  exclude companion flux. Please see Appendix A for more details on
  individual galaxies. A single vertical line marks the optical
  velocity for the galaxy (IC~1144) with an HI upper limit. IC~1144 is
  not in the sub-$M_b$ sample because of its large stellar mass ($M_*
  = 1.6\times10^{11}\,M_{\odot}$).\label{fig.gbtspectra_blue}}.
\end{figure*}

We check for companions to our GBT galaxies by setting the search
radius in the NASA/IPAC Extragalactic Database (NED) to match the beam
of the GBT. For galaxies with known or obvious companions, we measure
the HI flux twice: the first time we measure the flux within the
velocity range from ionized-gas rotation curves (or stellar rotation
curves, if ionized gas is not available), and the second time we
include all flux in the beam within a reasonable velocity range near
the galaxy ($\pm 300\, {\rm km\,s^{-1}}$). We mark the
ionized-gas/stellar velocity range used to measure fluxes for the
first method in Figures \ref{fig.gbtspectra_red} and
\ref{fig.gbtspectra_blue} with double vertical lines. We present both
measurements in Table \ref{table.newhi}: the first row contains values
measured in the ionized-gas/stellar velocity range, while the second
row includes companion flux (if any) within $\pm$300 ${\rm
  km\,s^{-1}}$.

The first method most likely underestimates HI content for the two
galaxies (IC~1639, IC~195) for which we must determine the velocity
range from stellar rotation curves, as those rotation curves are still
rising at the last measured point. The ionized-gas rotation curves
used for the other galaxies, on the other hand, are flat, so the
underestimation of HI flux for those galaxies should be small. The
second method likely overestimates a galaxy's HI reservoir, since cold
gas from a companion may not be readily accessible for star formation
in the target galaxy. {\it We use fluxes measured with only the first
  method in our analysis, as conservative estimates of HI gas mass.}
We discuss how our results may change if we include companion gas in
\S \ref{section.simulations}.

\subsubsection{Kinematics}
We estimate the observed maximum rotation speed, $V^{{\rm sin} i}_{\rm
M}$ for each galaxy following \citet{paturel03b}:

\begin{equation}
{\rm log} 2V^{{\rm sin} i}_{\rm M} = a\,{\rm log}W(r,l) + b
\label{equation.velo}
\end{equation}


\noindent 
where $a$ and $b$ are specified for a given velocity width $W$
measured at \% level $l$ and velocity resolution $r$, enabling
galaxies from disparate HI datasets to be compared. For our GBT data,
we adopt the recommended values of $a = 1.071$ and $b = -0.21$ to
convert our $W_{50}$ widths into $V^{{\rm sin} i}_{\rm M}$ (Table
\ref{table.nfgshi}, ~Col.~(7)).

We find the difference between heliocentric measurements of the HI
velocity and the optical velocity for NFGS galaxies to be small,
centered around 0 ${\rm km\,s^{-1}}$, with a standard deviation of 33
${\rm km\,s^{-1}}$.

\subsubsection{Atomic Gas Masses}
\begin{figure}
\includegraphics[scale=.7,angle=0]{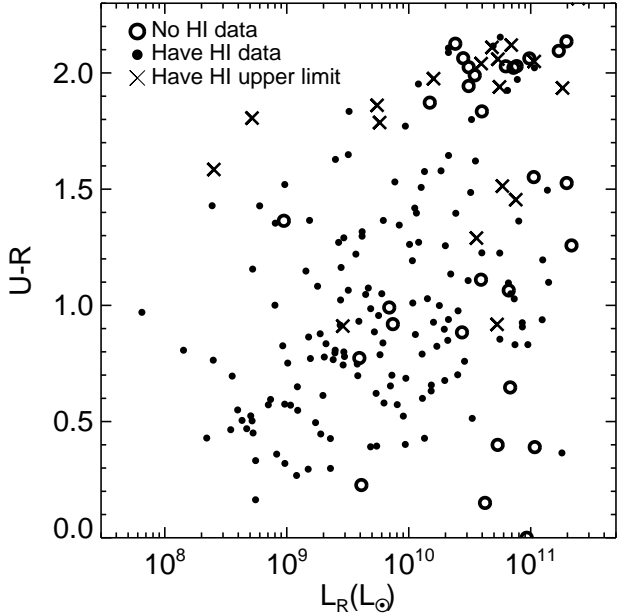}
\caption{$U-R$ color as a function of $R$-band luminosity, both
  extinction corrected, for galaxies in the NFGS. HI detections are
  shown as dots, HI upper limits are represented by $\times$'s, and
  galaxies with no HI data are shown as open
  circles. \label{fig.hisample}}
\end{figure}

Of the 200 galaxies in the NFGS, we have HI information for 170: new
GBT observations for 27 galaxies, HyperLeda HI data for 128 galaxies,
and other literature data for 15 galaxies. The 30 galaxies with no HI
information are distributed reasonably evenly between morphological
types (53\% early, 40\% late, and 7\% Pec/Im) as well as sequences
(43\% blue, 43\% red, and 14\% unknown because no stellar mass
estimate is available). We plot $U-R$ color as a function of $R$-band
luminosity in Figure \ref{fig.hisample} to show the distribution of
the NFGS sample with HI information. There is a somewhat higher
frequency of missing data among the brightest galaxies (also the most
distant, \citealt{jansen00a}).

The HI gas masses given in ~Col.~(9) of Table \ref{table.nfgshi} are
calculated from $f_{\rm HI}$ following \citet{haynes84}:

\begin{equation}
{M_{\rm HI} = 2.36\times10^{5}\,f_{\rm HI}\,(\frac{v_{\rm vlgvc}}{H_0})^2 \,\, {M_{\odot}}}
\label{equation.himass}
\end{equation}

\noindent where ${v_{\rm vlgvc}}$ is the Local Group and Virgocentric
flow corrected recessional velocity from \citet{jansen00a}, which we
use to be consistent with the stellar mass estimates. We multiply the
HI gas mass by a factor of 1.4 to account for the presence of helium.

\subsubsection{Upper Limits to HI Masses}
To calculate upper limits, we measure the rms dispersion of the
baseline, $\sigma_{\rm chan}$, in a signal-free part of the spectrum
within the velocity range from ionized-gas (or stellar) rotation
curves. We then estimate $\sigma_{\rm up}$ following:

\begin{equation}
\sigma_{\rm up} = \sigma_{\rm chan}\,\Delta V\, {\rm\sqrt{N} }
\end{equation}

\noindent where N is the number of channels in the velocity range we
measured from, and $\Delta V$ is the width of each channel in units of
${\rm km\,s^{-1}}$. We estimate the HI mass upper limit using
3${\sigma_{\rm up}}$ as the HI flux and following Equation
\ref{equation.himass}.

\section{Gas Reservoirs}\label{section.gasreservoirs}
\subsection{Comparison of Gas Reservoirs in NFGS Galaxies}
All of our sub-$M_b$ blue-sequence E/S0s, as well as both mid-sequence
E/S0s, are detected in HI and have atomic gas masses ($M_{\rm HI+He}$)
ranging from $10^7$ to almost $10^{10}\,{M_{\odot}}$. In contrast,
four of our eleven sub-$M_b$ red-sequence E/S0s were not detected,
although we integrated down to upper limits of $10^6$--$10^7\,
{M_{\odot}}$. The remaining seven galaxies have atomic gas masses
ranging from $10^7$--$10^9\,{M_{\odot}}$.

\begin{figure}
\includegraphics[scale=.7,angle=0]{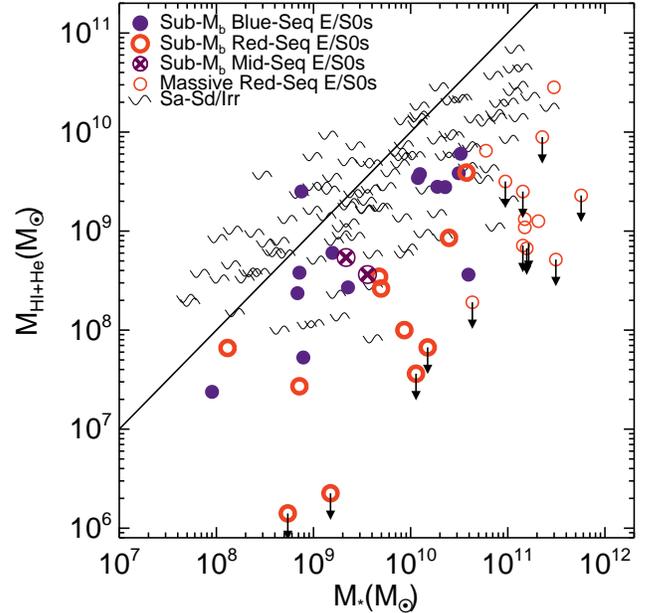}
\caption{Distribution of atomic gas mass (HI + He) for galaxies in the
  NFGS, as a function of stellar mass. Solid line indicates 1:1 ratio;
  downward arrows indicate upper limits. \label{fig.mgasvmstar}}
\end{figure}

We plot the distribution of atomic gas mass as a function of stellar
mass for the sub-$M_b$ blue- and red-sequence E/S0s, mid-sequence
E/S0s, and other NFGS galaxies in Figure \ref{fig.mgasvmstar}.  We
find that, at a given ${M_*}$, blue-sequence E/S0s tend to have larger
gas masses than red-sequence E/S0s. This result confirms the
preliminary results of \citetalias{kannappan09a}, for HI data that are
much more complete in terms of sampling sub-$M_b$ E/S0s. Since almost
half of the red-sequence E/S0s are actually upper limits, the
separation between the gas masses of blue- and red-sequence E/S0s is
actually larger than it appears in Figure \ref{fig.mgasvmstar}.

\begin{figure*}
\hspace{-2cm}
\epsscale{1.2}
\plottwo{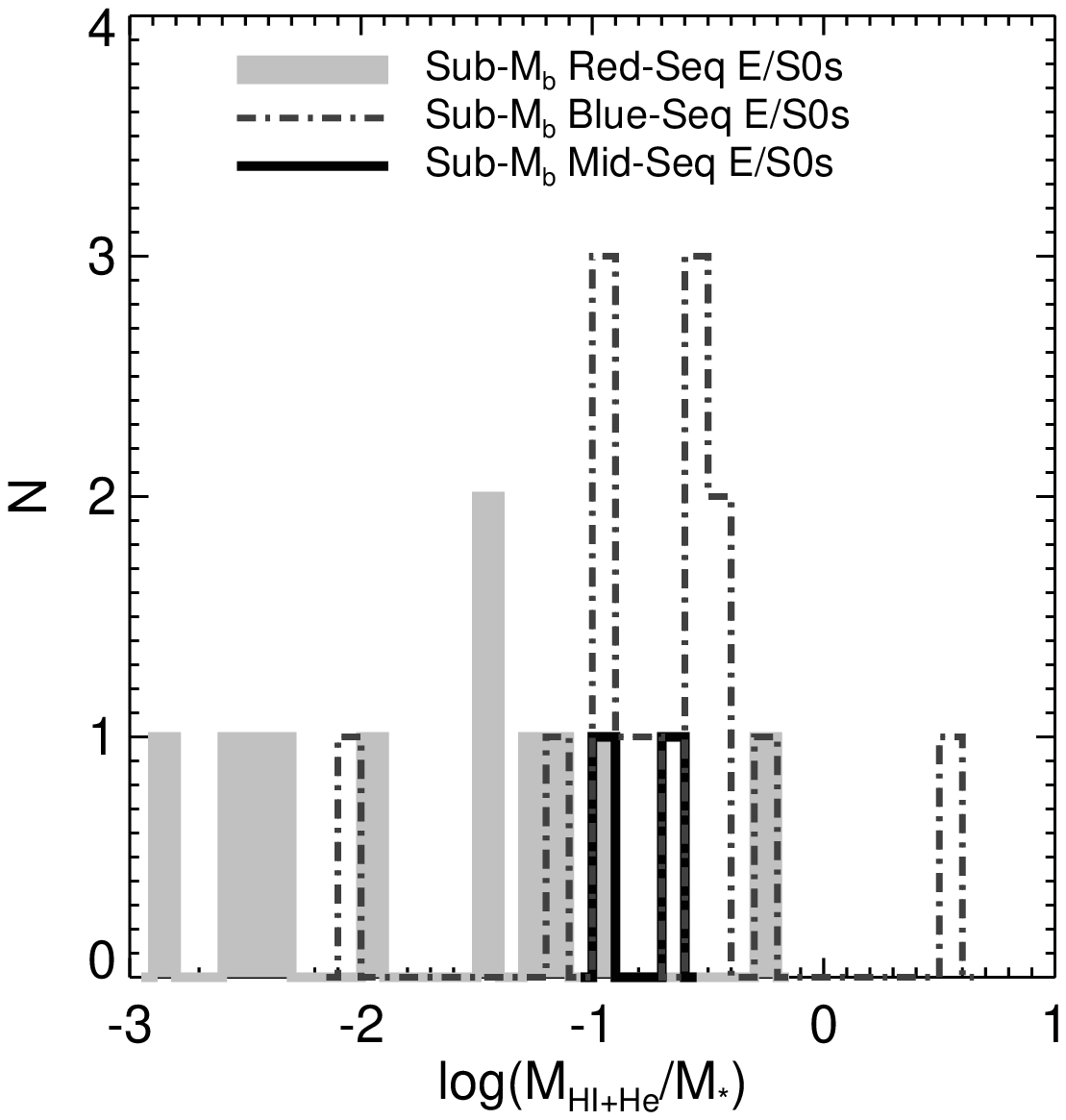}{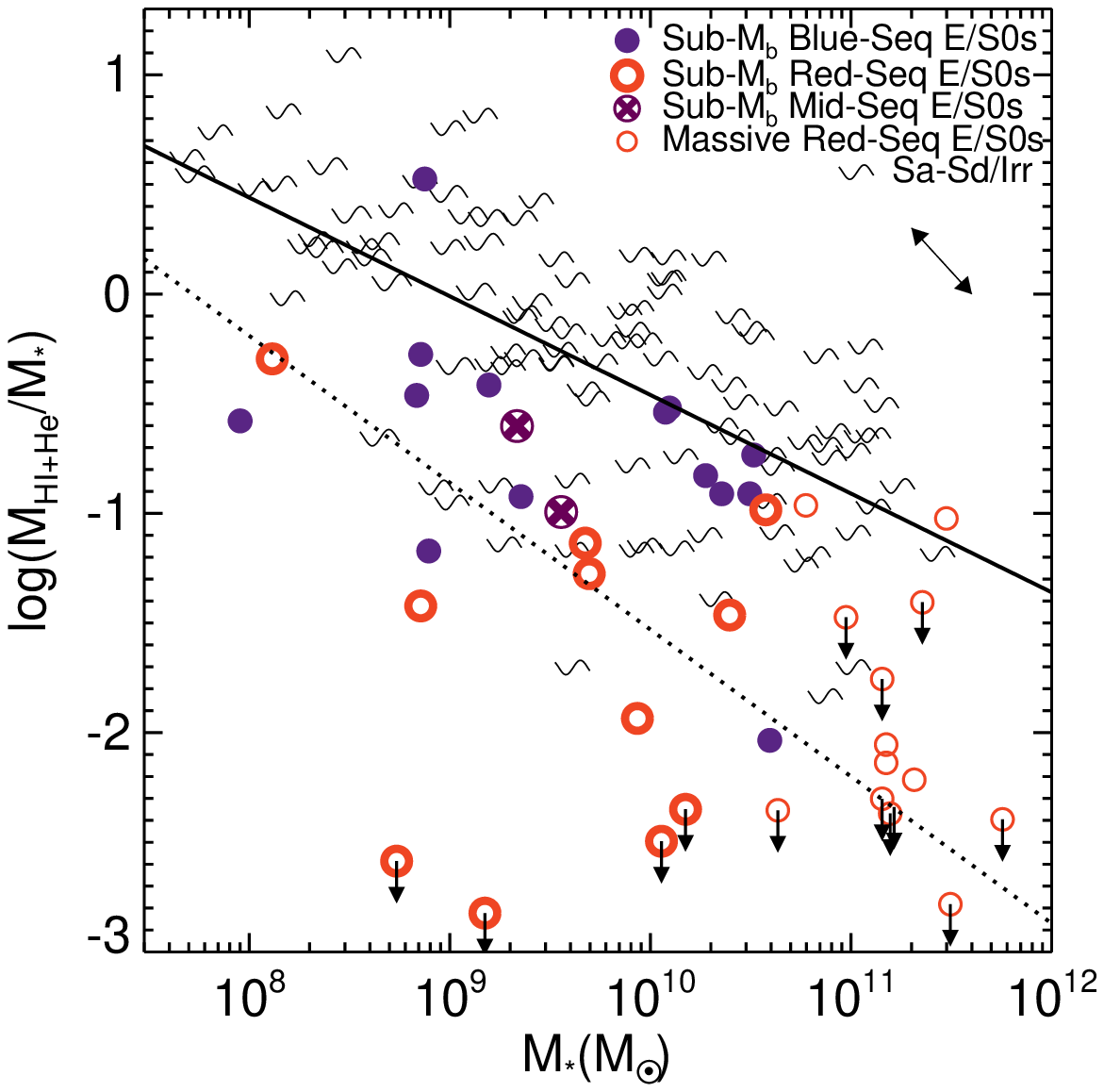}
\hspace{-1cm}
\caption{(a) Histogram of the distribution of atomic gas-to-stellar
  mass ratio for red-, blue-, and mid-sequence E/S0s in the
  sub-${M_b}$ E/S0 sample. (b) Atomic gas-to-stellar mass ratio as a
  function of stellar mass for galaxies in the NFGS. The arrow in the
  upper right indicates a factor of two error in stellar mass in
  either direction (increasing or decreasing). Downward arrows
  indicate upper limits. The solid/dashed line is the forward fit of
  ${M_{\rm HI+He}/M_*}$ as a function of ${M_*}$ for late/early-type
  galaxies in the NFGS. \label{fig.gsvmstar}}
\end{figure*}

Even more intriguing is the location of blue-sequence E/S0s {\it in
  between} spiral/irregular galaxies and red-sequence E/S0s in the
${M_{\rm HI+He}}$ vs. ${M_*}$ relation shown in Figure
\ref{fig.mgasvmstar}. In fact, there is considerable overlap between
blue-sequence E/S0s and spiral/irregular galaxies. This supports the
\citetalias{kannappan09a} suggestion that blue-sequence E/S0s form a
transitional class between spirals/irregular galaxies and traditional
red-sequence E/S0s, as originally inferred from the fact that
blue-sequence E/S0s also lie between spiral/irregular galaxies and
red-sequence E/S0s in the $M_*$-radius and $M_*$-$\sigma$ relations.

Since we are interested in the potential for morphological
transformation in blue-sequence E/S0s, it is informative to consider
${M_{\rm HI+He}/M_*}$ --- the mass of the atomic gas relative to the
current stellar mass of a galaxy. We plot the distribution of atomic
gas-to-stellar mass ratios for sub-$M_b$ E/S0s on both sequences in
Figure \ref{fig.gsvmstar}a and list their values in Table
\ref{table.masses}. A Kolmogorov-Smirnov test on the distributions of
${M_{\rm HI+He}/M_{*}}$ for sub-$M_b$ red- and blue-sequence E/S0s rejects
at the 99\% level the possibility that these galaxies derive from the
same parent population.

While most of the sub-$M_b$ blue-sequence E/S0s and both the
mid-sequence E/S0s have ${M_{\rm HI+He}/M_*}$ in the range of 0.1 to
1.0, all but two of the sub-$M_b$ red-sequence E/S0s have ${M_{\rm
    HI+He}/M_*} < 0.1$. Following \citet{binney98} Figure 4.51, the
formation of new stars in an extended disk constituting 25\% of
original total stellar mass will change the typical S0 galaxy to an Sa
galaxy.  This suggests that at least half of the sub-$M_b$
blue-sequence E/S0s \textit{do} have large enough gas reservoirs for
major morphological transformation \textit{if} all the gas is
converted into stars in the disk. We discuss in later sections whether
the gas is actually forming stars. But for now, we can say that at
least half of our sub-$M_b$ blue-sequence E/S0s do have the {\it
  potential} to transform their morphologies simply based on their
atomic gas masses. In contrast, sub-$M_b$ red-sequence E/S0s lack
sufficient atomic gas to do the same with rare exceptions.

\subsubsection{Interesting Outliers}\label{section.outliers}
Figure \ref{fig.gsvmstar}a shows that while sub-$M_b$ red-sequence
E/S0s tend to have lower values of ${M_{\rm HI+He}/M_*}$, sub-$M_b$
blue-sequence E/S0s tend to have higher values. There are, however, a
couple of outliers that do not seem to follow this trend. The most
prominent of the outliers are the red-sequence E/S0 with a large gas
reservoir, UGC~5923 (log~${M_{\rm HI+He}/M_*} = -0.3$), and the
blue-sequence E/S0 with a very small gas reservoir, IC~1639
(log~${M_{\rm HI+He}/M_*} = -2.0$). Taking the stellar masses of these
galaxies into consideration, however, provides plausible explanations
for their gas-to-stellar mass ratios.

UGC~5923, the gas-rich red-sequence E/S0, has the lowest stellar mass
of all red-sequence E/S0s in the NFGS with ${M_*} = 1.3\times\,10^8
{M_{\odot}}$, so it is not surprising that this galaxy has a
fractionally large gas reservoir, despite its red color. ${\rm
  H}\alpha$ emission indicates that there is some low-level star
formation in UGC~5923, but not enough to push the galaxy towards the
blue sequence. This galaxy does, however, appear dusty and has the
highest internal extinction of all sub-$M_b$ red-sequence E/S0s, which
suggests that it could also be forming stars {\it now} behind an
obscuring dust screen. But for dust, this object might well follow the
trend of the blue-sequence E/S0s in Figure \ref{fig.gsvmstar}.

IC~1639, the gas-poor blue-sequence E/S0, is on the other end of the
stellar mass scale as the galaxy with the largest stellar mass in the
sub-$M_b$ E/S0 sample ($M_* = 3.9 \times 10^{10}\,
M_{\odot}$). \citetalias{kannappan09a} argue that blue-sequence E/S0s
at these higher stellar masses are more often associated with violent
encounters than disk building. It is possible that this galaxy
underwent an interaction with its larger companion in the
not-so-distant past, which triggered a burst of star formation (hence
its blue color) and quickly exhausted its gas reservoir.

\subsubsection{${M_{\rm HI+He}/M_*}$ as a Function of Stellar Mass}
The outliers described in the previous section indicate the importance
of taking stellar mass into account when considering the
gas-to-stellar mass ratio. Hence we plot ${M_{\rm HI+He}/M_{*}}$ as a
function of ${M_*}$ (Figure \ref{fig.gsvmstar}b) for all galaxies
in the NFGS.

We fit the trend of decreasing atomic gas-to-stellar mass ratio with
increasing stellar mass in Figure \ref{fig.gsvmstar}b with a line in
the form of ${{\rm log}(M_{\rm HI+He}/M_*) = m\,{\rm log}(M_*) + b}$
for all NFGS galaxies with HI data, using survival analysis (the ASURV
package: \citealt{asurv}) to include galaxies with ${M_{\rm HI+He}}$
upper limits. Table \ref{table.gsfits} lists the coefficients for the
different fits (forward, bisector) for different populations (Blue
Sequence, Red Sequence, Spiral/Irregulars, and E/S0s).

As mentioned earlier, most of the sub-$M_b$ blue-sequence E/S0s have
${M_{\rm HI+He}/M_{*}}$ in the range of 0.1 to 1.0, while all but two of
the sub-$M_b$ red-sequence E/S0s have ${M_{\rm HI+He}/M_*} < 0.1$. The
separation between the two populations is even more striking in Figure
\ref{fig.gsvmstar}b, and re-emphasizes the difference in potential for
morphological transformation between sub-$M_b$ blue- and red-sequence
E/S0s.

\subsection{Comparison with Sage \& Welch E/S0 Sample}
We compare the masses of gas reservoirs in our sample of sub-$M_b$
E/S0s with an ongoing survey of cold gas (HI and CO) in a
volume-limited sample of nearby E/S0s by \citet{sage06} and
\citet{sage07}. \citet{sage06} report that E/S0s contain less than
$\sim$10\% of the cold gas mass (${\rm HI+He+H_2}$) predicted for gas
return by analytical stellar evolution models \citep{ciotti91}. In
particular, \citet{ciotti91} predict that ${M_{\rm gas} \sim L_B}$ in
solar units, while \citet{sage06} conclude that ${M_{\rm gas} \lesssim
  0.1\, L_B}$ for S0s and suggest that this also applies to
ellipticals based on preliminary data \citep{sage07}.

\begin{figure}[!h]
\includegraphics[scale=0.7,angle=0]{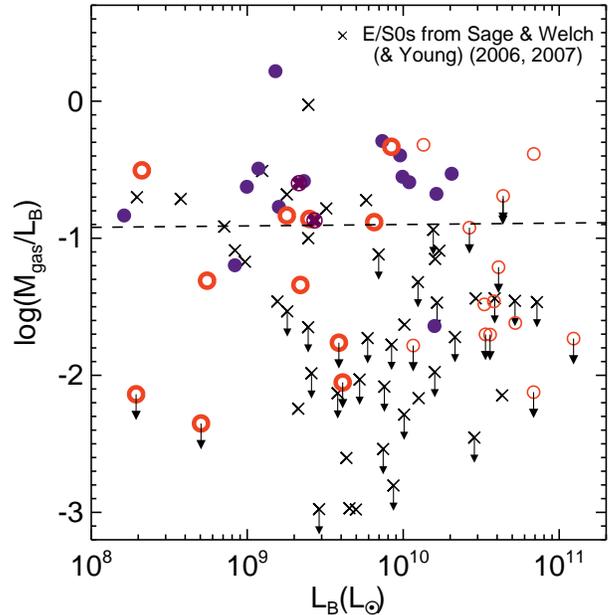}
\caption{Ratio of gas mass to $B$-band luminosity (in solar units)
  vs. $B$-band luminosity. The $B$-band luminosities are corrected for
  foreground and internal extinction. ${M_{\rm gas}}$ values for E/S0s
  from Sage \& Welch (2006, 2007; $\times$'s) account for HI, He, and
  ${\rm H_2}$, while ${M_{\rm gas}}$ values for blue-, red-, and
  mid-sequence E/S0s from this paper (filled and empty circles, same
  symbols as Figures \ref{fig.mgasvmstar} and \ref{fig.gsvmstar}b)
  account for only HI and He. The cold gas cutoff at ${M_{\rm gas}
    \sim 0.1\, L_B}$ found by \citet{sage06} is shown as a dashed
  line. Downward arrows indicate upper
  limits.\label{fig.blumvgasfrac}}
\end{figure}

We plot ${M_{\rm gas}/L_B}$ vs. ${L_B}$ in Figure
\ref{fig.blumvgasfrac}, showing ${M_{\rm gas}} \sim0.1\, {L_B}$ as a
dashed line. We divide ${M_{\rm gas}}$ by ${L_{B}}$ rather than $M_*$
to facilitate comparison with \citet{sage06}; although the larger
scatter compared to Figure \ref{fig.gsvmstar}b suggests that ${L_B}$
does not work as well as ${M_*}$, ${M_*}$ is not available for the
E/S0s from \citet{sage06} and \citet{sage07}.

While there are a few E/S0 galaxies from \citet{sage06} and
\citet{sage07} with values of ${M_{\rm gas}/L_B} >$ 0.1, the majority
of their E/S0s have much smaller values of ${M_{\rm gas}/L_B}$. In
contrast, all but two of our sub-$M_b$ blue-sequence E/S0s fall above
the 0.1~${M_{\rm gas}/L_B}$ cutoff, as do both of the mid-sequence and
a few red-sequence E/S0s. Note that our data points do not include
molecular gas as opposed to the \citet{sage06} data points, so the
values of ${M_{\rm gas}/L_B}$ for our galaxies reflect only the
neutral atomic ISM, and the actual values could be even higher. This
result suggests that the sub-$M_b$ blue-sequence E/S0s in our sample
have some of the most massive fractional gas reservoirs among
early-type galaxies, in comparison to red-sequence E/S0s in our sample
as well as E/S0s from the literature.

The differences in how galaxies in the two samples are selected may
explain the disparity in ${M_{\rm gas}/L_B}$ between them. The
\citet{sage06} sample, selected from the Nearby Galaxies Catalog,
inherits the parent sample's biases against optically small galaxies
(diameter $< 1\arcmin.5$--$2\arcmin$) and HI-poor systems
\citep{welch03}. Hence, it is likely that the \citet{sage06} E/S0s are
more massive and therefore more likely to be on the red sequence than
our sample of E/S0s. The distribution of ${L_B}$ for the two different
samples in Figure \ref{fig.blumvgasfrac} hints at this, with our
sample of sub-$M_b$ E/S0s predominantly at lower ${L_B}$ and the
\citet{sage06} sample dominating at higher values. Note that the
${L_B}$ of blue-sequence E/S0s in our sub-$M_b$ sample may be
systematically enhanced by star formation, decreasing ${M_{\rm
    gas}/L_B}$, so the difference in $M_{\rm gas}/M_*$ between our
sub-$M_b$ E/S0s and the \citet{sage06} galaxies may be even larger.

Although our sub-$M_b$ blue-sequence E/S0s have larger gas mass
reservoirs than seen by \citeauthor{sage06}, we still find much
smaller reservoirs than that predicted for gas return by the
analytical stellar evolution models discussed in \citet{sage06},
ranging from 10\% to 30\% of the predicted value. Our findings support
Sage \& Welch's conclusion that stellar mass loss is not the primary
source of HI in E/S0s.

\section{Star Formation and Stellar Mass Growth}\label{section.starformation}
Now that we have established that sub-$M_b$ blue-sequence E/S0s
typically have substantial fractional atomic gas reservoirs and
therefore potential for morphological transformation, we consider the
question of whether this gas is being converted into stars at a rate
that can lead to morphological transformation in a reasonable amount
of time. Although we lack information about the spatial distribution
of the atomic gas, we can make some simplifying assumptions and create
limiting scenarios for the evolutionary trajectory of blue-sequence
E/S0s given the current rate of star formation.

\subsection{Two Limiting Scenarios for Growth in the Stellar Component}\label{section.scenarios}
To construct truly realistic scenarios for the evolutionary path of
our galaxies, we would have to account for all sinks and sources of
gas. The sink terms are the rates at which gas is converted into stars
and ionized and/or expelled due to stellar winds and supernovae. The
source terms include the rate at which fresh gas is brought in from
external sources (minor mergers, interactions, etc.)  and at which gas
is returned by stellar evolution. A detailed accounting of all these
processes, however, is beyond the scope of this paper. While there are
many theoretical studies of the hierarchical assembly of galaxies
(many of which include estimates of the frequency of mergers),
simulations still lack the resolution to predict the frequency of very
minor mergers and interactions. We discuss current simulations in more
detail in \S \ref{section.simulations}.

We consider here the range of plausible evolutionary trajectories for
our blue-sequence E/S0s by presenting two simplified, but limiting,
scenarios. We should note that we assume that the atomic gas is
distributed in such a way that it can be made available for star
formation (e.g., via conversion to molecular gas) at a rate comparable
to the current star formation rate. This may not always be the case,
but we argue that this is a plausible assumption in \S
\ref{section.simulations} and \S \ref{section.inflow}.

In the first scenario, we assume that the current star formation rate
remains constant over time, which is an assumption many population
synthesis models make. Preserving a constant star formation rate
requires an increasing star formation efficiency for a closed box, or
open-box inflows. This scenario provides a reasonable upper limit on
the possible growth in stellar mass per unit time and a lower limit on
the amount of time it takes for that mass to form.

The second scenario represents the other limit: an exponentially
declining star formation rate (i.e., closed box with no gas
return). For each galaxy, we start out with the current cold gas
reservoir (${M_{\rm HI+He,0}}$) and star formation rate (${\rm
SFR_0}$). As time progresses, the star formation rate declines
exponentially following ${{\rm SFR}(t) = {\rm SFR_0}\, e^{-t/\tau}}$
as gas is converted into stars, where ${\tau = M_{\rm HI+He,0}/{\rm
SFR_0}}$ (e.g., \citealt{li06}). This scenario defines the lower
limit on the amount of gas converted to stars within a fixed time, as
the star formation rate is declining exponentially.

Because we lack information regarding the frequency of
internal/external gas replenishment, these two scenarios
(exponentially declining SFR with no new gas, and constant SFR
allowing gas infall) represent simplified limits which likely bracket
the actual amount of growth in the stellar component in these
galaxies.

\subsection{Estimates of Stellar Mass Growth Over Time}\label{section.massgrowth}
\begin{figure*}
\hspace{-.5cm}
\epsscale{1.1}
\plottwo{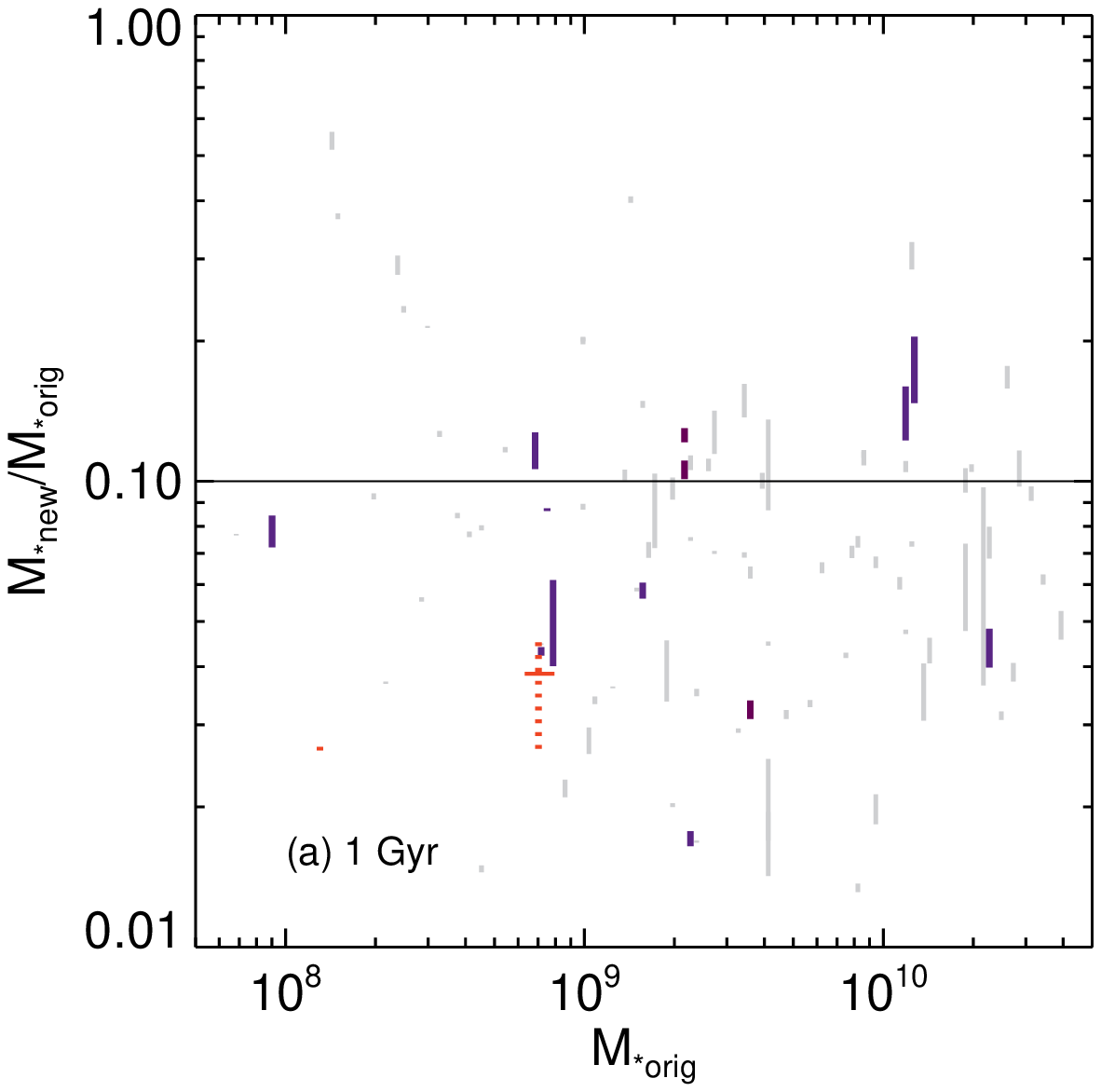}{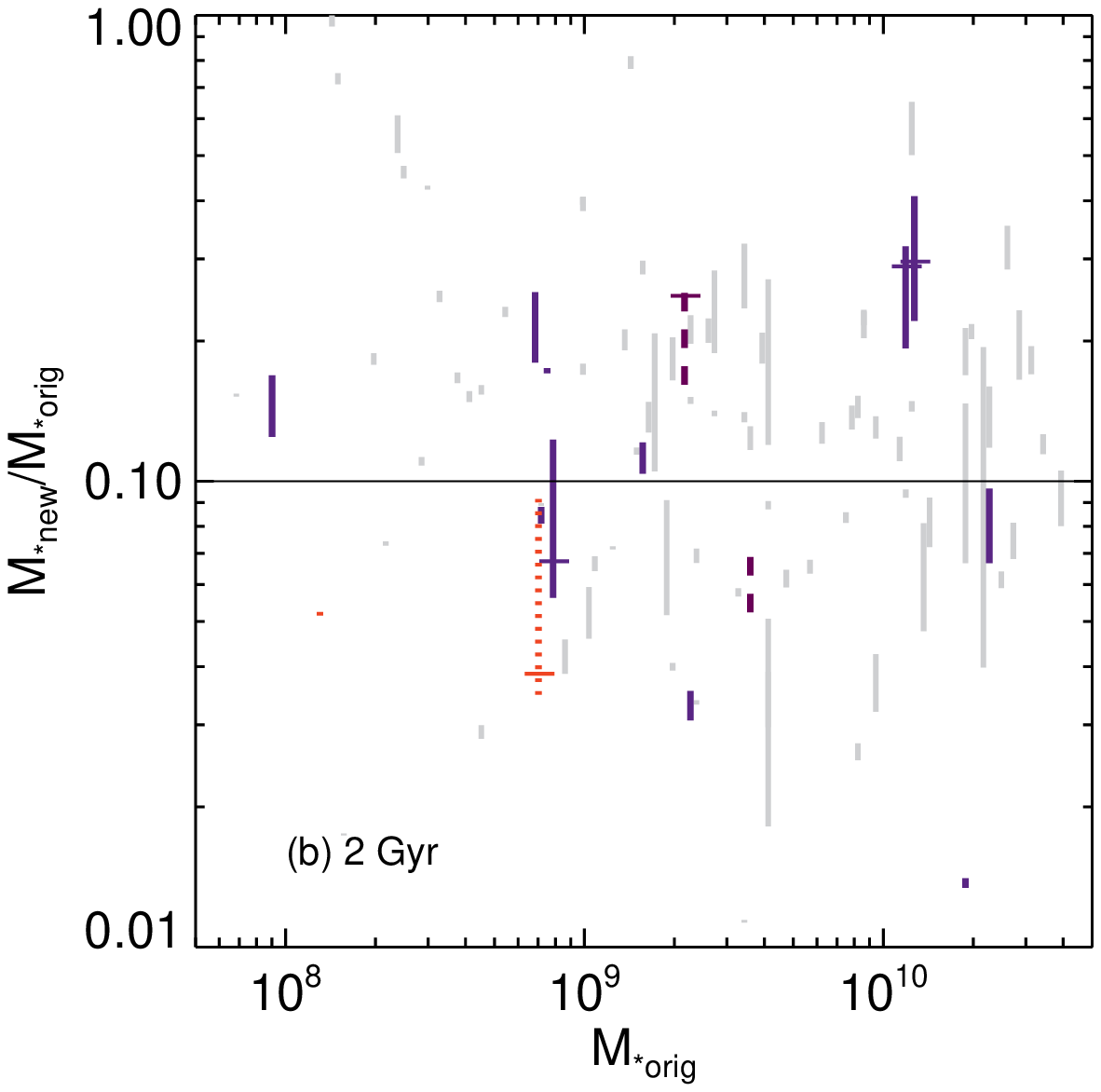}
\hspace{-1cm}
\plottwo{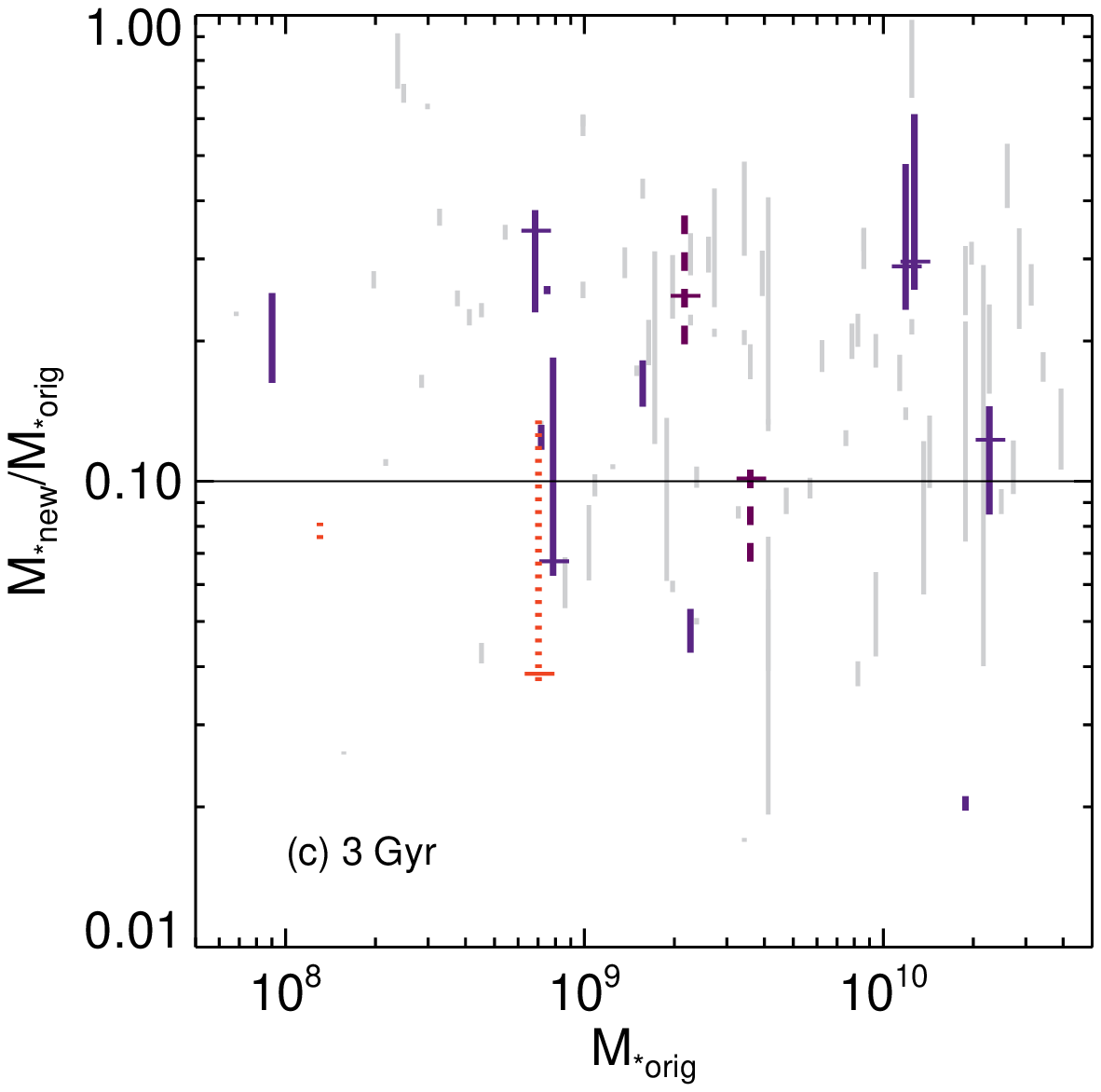}{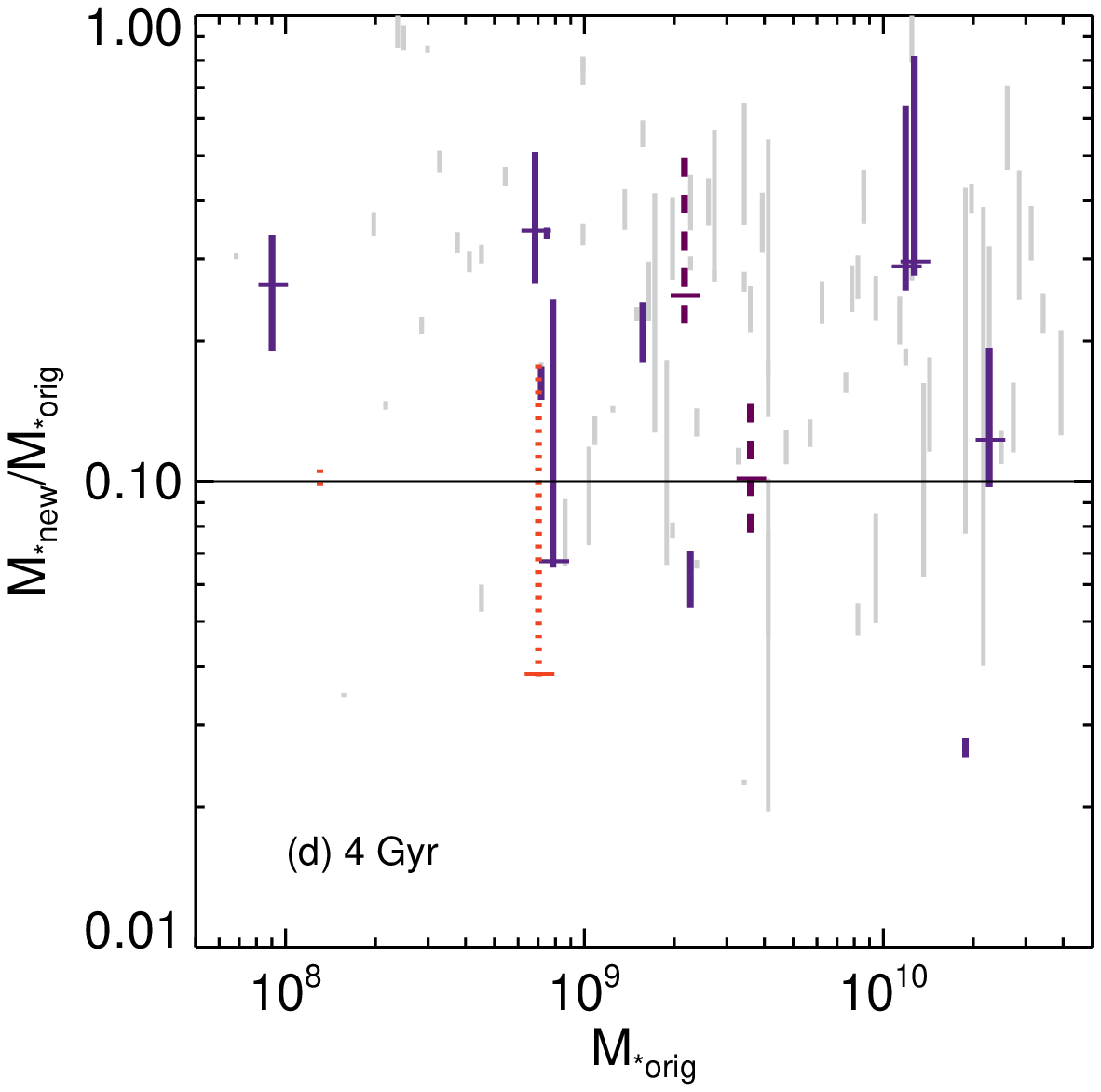}
\caption{The fractional stellar mass formed 1, 2, 3, and 4 Gyr in the
  future. For each galaxy, the two different scenarios are represented
  by the lower (exponentially declining SFR) and upper (constant SFR)
  end of a vertical line. The horizontal dash marks where the original
  gas reservoir runs out for each galaxy. The solid horizontal line
  indicates 10\% fractional stellar mass growth. Dark solid vertical
  lines represent sub-$M_b$ blue-sequence E/S0s, short vertical dashes
  are sub-$M_b$ red-sequence E/S0s, long vertical dashes are
  mid-sequence E/S0s, and solid grey lines are spiral/irregular
  galaxies.\label{fig.newstarsformed}}
\end{figure*}

Given the two limiting scenarios described in the previous section,
how much growth in stellar mass is possible for the different
galaxies? We plot the ratio of new stellar mass formed relative to the
current stellar mass as a function of current stellar mass 1, 2, 3,
and 4 Gyr in the future in Figures \ref{fig.newstarsformed}a--d,
respectively. Here we note that 11 E/S0s from the sub-$M_b$ sample
(two on the blue sequence and nine on the red sequence) do not appear
in Figure \ref{fig.newstarsformed}. These galaxies have spectra that
are integrated over the galaxy, but no H$\alpha$ emission was
detected, so these galaxies do not appear in any of the figures using
star formation rates. Future work with {\it GALEX} and {\it Spitzer}
data will provide better estimates of the star formation rates in
these galaxies.

For each galaxy, a vertical line in Figure \ref{fig.newstarsformed}
represents the range of possible fractional increase in stellar mass,
with the lower end representing the exponentially declining SFR
scenario and the upper end representing the constant SFR scenario. We
mark the vertical line with a horizontal dash to note the stellar mass
fraction at which the original gas reservoir runs out. Any growth in
the stellar component indicated by the line above the horizontal dash
requires inflow of gas, which we allow in the constant star formation
scenario. Galaxies that have not converted all of the original gas
mass in the constant star formation rate scenario after the time
specified for each figure do not have a horizontal dash. Since these
two scenarios are the limiting cases, it is likely that the actual
amount of stellar mass growth is somewhere in between the two ends of
the line.

Figure \ref{fig.newstarsformed}b shows that many of the sub-${M_b}$
blue-sequence E/S0s (dark solid lines) have the potential to increase
their stellar masses by a large fraction in just 2 Gyr, with seven of
the fourteen sub-$M_b$ blue-sequence E/S0s crossing or above the 10\%
line. In 3 Gyr, the time by which most sub-$M_b$ blue-sequence E/S0s
have exhausted their gas reservoirs (\ref{section.timescale}), nine of
the fourteen sub-$M_b$ blue-sequence E/S0s will cross or be above the
10\% line. The remaining five blue-sequence E/S0s have lower SFRs, so
they are below the 10\% line after 3 Gyr has passed. In fact, the SFRs
for NGC~7360, IC~1639, and IC~195 are so low they do not form $>$0.1\%
of their current stellar masses within 3 or even 4 Gyr, so they do not
appear on any of the panels in Figure \ref{fig.newstarsformed}. This
large spread in the fractional stellar mass growth in blue-sequence
E/S0s reflects the spread in star formation rates, which in turn may
be indicative of differences in burst stages of the galaxies within
this population. We discuss this in more depth in \S
\ref{section.inflow}.

The two mid-sequence E/S0s in our sample also have potential for
substantial morphological transformation. UGC~7020A (${M_* \sim 2.2
  \times 10^9}$) can form new stellar mass in the range of 20--40\% of
its current stellar mass within 3 Gyr, and UGC~6570 (${M_* \sim 3.6
  \times 10^9}$) can form new stellar mass $\sim$10\% over the same
period.

The low-mass red-sequence E/S0 with a surprisingly large gas mass
reservoir discussed earlier, UGC~5923 (${M_* \sim 1.3 \times 10^8}$),
has a relatively low star formation rate, but can still form new
stellar mass $>$10\% of its current stellar mass within 3
Gyr. NGC~5338 (${M_* \sim 7.2 \times 10^8}$) has a very small gas
reservoir but might form significant stellar mass if there is
replenishment of gas. The rest of the sub-$M_b$ red-sequence E/S0s
have such low SFRs that they are below the 1\% line (i.e., off the
plot) in all four panels of Figure \ref{fig.newstarsformed}.

We plot the same lines of fractional stellar mass growth for spiral
and irregular galaxies with ${M_* \leq 4 \times 10^{10} M_{\odot}}$ in
grey in Figures \ref{fig.newstarsformed}a--d for comparison. For
sub-$M_b$ blue-sequence E/S0s, the ratio of new stellar mass formed in
1--3 Gyr to original stellar mass is comparable to that of the
spiral/irregular distribution (Figure
\ref{fig.newstarsformed}a--c). This suggests that, as a population,
sub-$M_b$ blue-sequence E/S0s have potential for growth in the stellar
component similar to that of spiral/irregular galaxies in the near
term future ($\sim$1--3 Gyr). The horizontal dashes marking where the
original gas reservoirs runs out, however, suggest that the inflow of
fresh gas is important to the long-term evolution of our galaxies
(Figure \ref{fig.newstarsformed}c, d).

\section{Availability of Gas for Star Formation}\label{section.sfmechanism}
\subsection{Timescale for Gas Exhaustion and Inflow}
Without maps of the distribution of HI, we cannot say for certain that
the atomic gas is distributed in such a fashion that it is readily
available for star formation. In \S \ref{section.timescale} and \S
\ref{section.simulations} below, we argue that regardless of the
distribution of the atomic gas (extra-planar, in the disk, or in
companions), the timescale for inward travel of gas is most likely
shorter than the duration of star formation in blue-sequence
E/S0s. This suggests that assuming that the atomic gas reservoir is
readily available for star formation is reasonable given internal or
external mechanisms to drive gas inflow. In \S \ref{section.inflow} we
will discuss evidence for frequent, externally driven inflow in the
sub-$M_b$ blue-sequence E/S0 population.

\subsubsection{Gas Exhaustion Time}\label{section.timescale}
The timescale that can be directly estimated from the atomic gas mass
and star formation rate of a galaxy is the gas exhaustion time ($\tau
= {M_{\rm HI+He}/{\rm SFR}}$) --- the amount of time it would take to
convert all the gas into stars, assuming the current star formation
rate remains constant. This is the same $\tau$ as the one used in the
exponentially declining star formation rate calculation for the second
scenario in \S \ref{section.scenarios}, although by definition the gas
reservoir will never be exhausted in this scenario since the SFR
decreases exponentially in parallel with decreasing gas mass. {\it
  Most} of the gas mass, however, will be converted into stellar mass
within the timescale $\tau$, so this is an interesting timescale to
consider for both scenarios.

Table \ref{table.gastimes} lists the gas exhaustion times for galaxies
in the sub-$M_b$ E/S0 sample that have star formation rates. The large
spread in gas exhaustion times may be reflective of the diversity of
evolutionary states within the blue-sequence E/S0 population, which we
explore further in \S \ref{section.inflow}. Note that the range of gas
exhaustion timescales we find includes shorter timescales than found
by KGB because of preferential incompleteness at low HI masses in the
archival data used by KGB.

\begin{figure}[!h]
\includegraphics[scale=.7,angle=0]{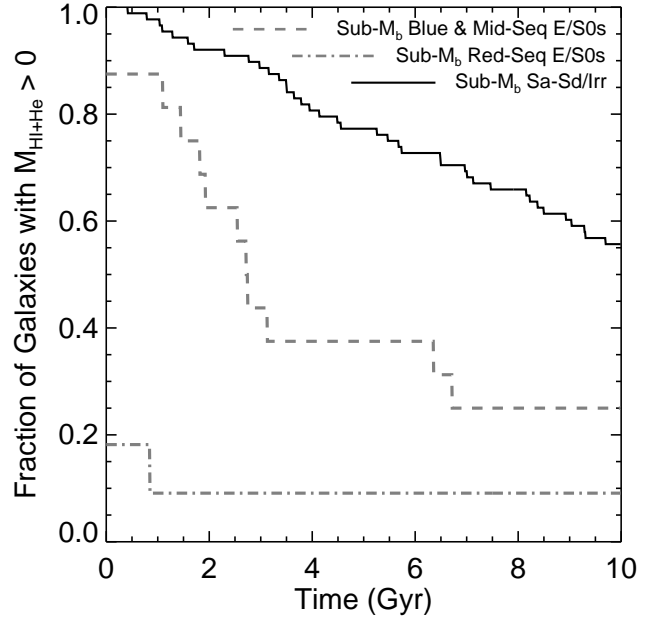}
\caption{Fraction of galaxies that have not exhausted their atomic gas
  reservoir as a function of time in a constant star formation rate
  scenario. Fractions begin below 1.0 because some E/S0s have no
  detected star formation.\label{fig.timescales}}
\end{figure}

For a more informative picture, we plot the fraction of sub-$M_b$
galaxies that will {\it not} have exhausted their original atomic gas
reservoirs at a given future time in Figure \ref{fig.timescales}. We
include the mid-sequence E/S0s with the blue-sequence E/S0s in this
figure as they seem to behave similarly. While the fraction of
star-forming spiral/irregular galaxies seems to have a gradual, smooth
decline in this figure, the blue-sequence E/S0s have a sharp drop-off
at $\sim$3 Gyr. In agreement with Figure \ref{fig.newstarsformed}a--d,
Figure \ref{fig.timescales} suggests that sub-$M_b$ spiral/irregular
galaxies typically continue forming new stars long after star
formation in sub-$M_b$ blue-sequence E/S0s is extinguished by the
exhaustion of the original gas reservoir.

There is no drop-off for the sub-$M_b$ red-sequence E/S0s, as we have
SFRs for only two of the eleven galaxies. The fact that most of the
sub-$M_b$ red-sequence E/S0s have extremely low star formation rates,
however, supports the conclusion that the near-term evolutionary
trajectories of blue- and red-sequence E/S0s will be quite different.

\subsubsection{Inward Travel of Gas}\label{section.simulations} 
In \S \ref{section.gasreservoirs} and \S \ref{section.starformation},
we have demonstrated that blue-sequence E/S0s have substantial
fractional atomic gas reservoirs that, if readily available for star
formation, can translate into significant growth in stellar mass and
consequent morphological transformation. The key uncertainty is
whether this gas is or can be made available for star formation on a
reasonable timescale compared to the gas exhaustion timescale just
discussed.

The GBT spectra, at 21~cm, have a resolution of only $\sim$9$\arcmin$,
which (for our sample) translates to an uncertainty radius of 10--110
kpc in the location of atomic gas, depending on the distance to the
galaxy. Without maps with sufficient angular resolution (e.g., VLA HI
maps), we do not know whether the cold atomic gas is accessible for
conversion into ${\rm H_2}$ for star formation or not. Broadly, there
are three possible distributions for the HI gas: on a trajectory
falling into the galaxy, in an HI disk that is a part of the galaxy
(the most plausible configuration), or in companions.

{\it Infalling Gas:} If the gas is somewhere outside the galaxy on its
way inward, we expect it to travel inwards to the galaxy on a
dynamical timescale. The 9$\arcmin$ GBT beam at 1.4 GHz corresponds to
radii of 10--110 kpc from the centers of galaxies in the sub-${M_b}$
E/S0 sample, with a median of 34 kpc. We list the dynamical timescale
for inward travel of gas from the edge of the beam for each galaxy in
the sub-$M_b$ sample in ~Col.~(5) of Table \ref{table.gastimes}. The
dynamical timescale estimates for inward travel of gas range from 0.4
to 2.5 Gyr, with a median of 1.1 Gyr. These timescale estimates are
smaller than the gas exhaustion times for all but two of the
galaxies. These two galaxies happen to be the most distant of all
sub-$M_b$ E/S0s in Table \ref{table.gastimes}, so the edge of the GBT
beam corresponds to $>$100 kpc. Therefore it is not surprising that
these galaxies have long timescales for infall from the edge of the
beam that are greater than the gas exhaustion time.

Assuming that the gas is at the edge of the beam is the most extreme
case; it is much more likely that the gas is much closer to the
galaxy, which would reduce the infall time as $r^{3/2}$. For example,
the dynamical time for inflow of gas from the predicted HI radius
(2--17 kpc, scaling from the blue optical radius using an assumed ratio
of 2.11; \citealt{noordermeer05}) is much shorter than the gas
exhaustion time in all cases, ranging from 50--400 Myr, with a median
of 150 Myr (Table \ref{table.gastimes}, ~Col.~(4)). \citet{fraternali08}
find evidence for the infall of extra-planar gas onto star-forming
spiral galaxies on short timescales --- at rates comparable to their
star formation rates ($\sim$few $M_\odot\,{\rm yr^{-1}}$), which
supports our estimates above.

The dynamical timescales we estimate here are for inward travel of gas
all the way to the center of the galaxy. However, the gas is capable
of forming stars far from the centers of galaxies, depending on
parameters such as local surface density and mid-plane pressure (e.g.,
\citealt{blitz04,leroy08}). Thus the distance infalling gas has to
travel to reach star forming regions and the corresponding infall
timescale may be even shorter than our estimates above.

{\it Disk Gas:} Comparison between ionized-gas rotation curves and the
new GBT HI profiles suggests that most of the atomic gas is likely
distributed in rotating disks for all the GBT galaxies. We identify
four particularly interesting cases. The ionized gas data show
marginal or no rotation for UGC~6805, UGC~7020A, and UGC~6003, but
their HI profiles have the appropriate widths for rotation given their
stellar masses (e.g., based on the $M_*$-rotation velocity relation in
\citetalias{kannappan09a}). There is no ionized gas detection at all
for NGC~3522, so the HI profile for this galaxy indicates the presence
of a previously unknown gas disk.

If the gas is in a stable orbit in the disk of a galaxy, it will not
necessarily travel inwards towards star forming regions on a short
timescale compared to the gas exhaustion time. Depending on the
density of the gas, the gas disk may or may not collapse to form
stars. A dense disk of gas could dovetail nicely with either of our
star formation scenarios above --- it will collapse and form a stellar
disk at the rate of the global SFR. If, however, the HI disk is too
diffuse and spread out in an extended disk, the gas will continue in
circular orbit in a dormant fashion unless it is perturbed by internal
instabilities or events such as minor mergers or interactions.

Recent simulations find that the frequency of mergers increases as the
ratio of masses between the progenitors ($\xi < 1$) decreases (e.g.,
\citealt{stewart08,fakhouri08} and references therein). At the finest
resolution currently available with the Millennium Simulation,
\citet{fakhouri08} find minor merger rates of 0.2 to 0.7 mergers per
halo per Gyr at z = 0 for $\xi =$ 1:30 to 1:100, respectively. These
minor merger rates correspond to one merger every 5 Gyr on the high
mass end (1:30) and one per 1.4 Gyr on the low mass end (1:100). The
minor merger rate at the high mass end (1:30) is perhaps a bit long
relative to the gas exhaustion times for our galaxies, but minor
mergers with progenitor mass ratios down to 1:100 and even smaller may
still be capable of inducing gas inflow and star formation. Due to the
lack of resolution, however, \citet{fakhouri08} do not consider
progenitor mass ratios smaller than 1:100, where mergers are
extrapolated to occur on timescales shorter than 1 Gyr.

There is also observational evidence suggesting that tidal
interactions with small companions occur relatively frequently in
field galaxies, bringing fresh infall of gas. At least 25\% of field
galaxies observed in HI in several surveys show asymmetric features,
indicating that they have recently undergone or are undergoing tidal
interactions \citep{sancisi92,verheijen01,vanderhulst01}. Moreover, if
one takes lopsided structure (azimuthal distortions in the stellar
disk) and kinematics as evidence of interaction, the fraction would
increase to above 50\% of field galaxies \citep{zaritsky97}.

The frequency of minor mergers, taking into account recent studies of
how gaseous disks could survive mergers and interactions
\citep{hopkins09,stewart09}, suggests that if some of the galaxies in
our sub-$M_b$ E/S0 sample have large, diffuse, and extended HI disks,
they may not lie dormant for too long before a minor merger or
interaction induces the gas to flow inwards and triggers star
formation. Once a minor merger or interaction occurs, we expect the
gas to travel inwards on the order of a dynamical timescale or shorter
\citep{barnes01}. As we discussed above, these are short timescales
relative to the frequency of minor mergers in simulations, and so we
take the merger rate as the limiting factor in this case, not the
subsequent inward travel time.

Here we have considered only external mechanisms that could drive the
disk gas inwards, but secular mechanisms that are internally driven
(e.g., cloud-cloud collisions that provide an effective viscosity,
bars that may be able to form independently of external perturbations,
resonances, instabilities, etc.) likely also play a role in the inflow
of disk gas. Because these mechanisms do not require external
triggers, including the effects of internal mechanisms will shorten
the gas inflow timescales we estimate from minor mergers/interactions
alone.

{\it Companion Gas:} As described in \S \ref{section.companions}, we
adopt a conservative approach and limit the HI flux measurement to the
velocity range indicated by the primary ionized-gas (or stellar)
rotation curve for galaxies with known companions. Since we restrict
the flux measurements, our fluxes are most likely underestimates.

Of the four sub-${M_b}$ E/S0s with known companions in NED
(UGC~12265N, IC~195, IC~1639, and NGC~4117), only UGC~12265N has a
non-negligible SFR and appears in the plots in \S
\ref{section.starformation}. The other three will increase their
atomic gas-to-stellar mass ratios by factors of 1.6--3.7 if we include
the companion gas (IC~195: 0.12 $\rightarrow$ 0.20, IC~1639: 0.01
$\rightarrow$ 0.03, NGC~4117: 0.05 $\rightarrow$ 0.12), but do not
change the results of \S \ref{section.gasreservoirs}
significantly. Because these three galaxies have no detected star
formation, the fractional stellar mass growth over time remains
negligible.

Including the companion gas, the atomic gas-to-stellar mass ratio for
UGC~12265N would quadruple from 0.3 to 1.2, giving it the second
highest value of ${M_{\rm HI+He}/M_*}$ for sub-${M_b}$ blue-sequence
E/S0s. The amount of growth over time (Figure
\ref{fig.newstarsformed}) would not change by much; the upper limit
would remain the same, but the lower limit would increase since $\tau$
is larger with the extra gas. If the atomic gas is readily accessible
for star formation, the gas exhaustion timescale would then increase
from 1.4 Gyr to 5.6 Gyr.

Following Equation 4 of \citet{lin83}, we estimate the timescale for a
merger via dynamical friction between UGC~12265N and its companion
(UGC~12265S) 12 kpc away (neglecting the unknown line-of-sight
distance) to be $\sim$70 Myr \footnote{ $t_{merge} = \frac{10^{10}
    {\rm yr}}{ln \Lambda}\, [\frac{r_s(0)}{52 {\rm kpc}}]^2 \,
  [\frac{10^{10}\,M_{\odot}}{m_s}]\, [\frac{v_c}{\sqrt(2)\cdot 100
    {\rm km\,s^{-1}}}] $. We assume that ln $\Lambda$ = 3.3, $v_c =
  250 {\rm km\,s^{-1}}$, and the companion is half as massive as
  UGC~12265N. If the companion is more massive than our assumption,
  the merger time would decrease by the same factor. If the companion
  is less massive and/or further away, the timescale for merger would
  increase as $m_s^{-1}$ and $r_s^2$.}. This is shorter than its gas
exhaustion timescale, which suggests that much of the companion gas is
available for star formation in the near future and the amount of
growth in the stellar component for this galaxy is greater than our
conservative estimates in the previous sections. 

Given the 1:2 progenitor mass ratio for this system, the resultant
burst of star formation is likely to be more extreme than the
scenarios discussed in \S\ref{section.scenarios}, enhancing the star
formation rate by a factor of two or more (e.g.,
\citealt{li08,darg09}). Whether the merger between UGC~12265N and its
companion and the subsequent star formation will result in late-type
morphology is unclear. A burst of central star formation may leave
UGC~12265N with early-type morphology (e.g., \citealt{dasyra06}), but
the gas-richness of this pair suggests that a disk-dominated remnant
may be more likely (\citetalias{kannappan09a};
\citealt{hopkins09,stewart09}).

\subsection{Evidence for Episodic Gas Inflow}\label{section.inflow}

\begin{figure}
\includegraphics[scale=.7,angle=0]{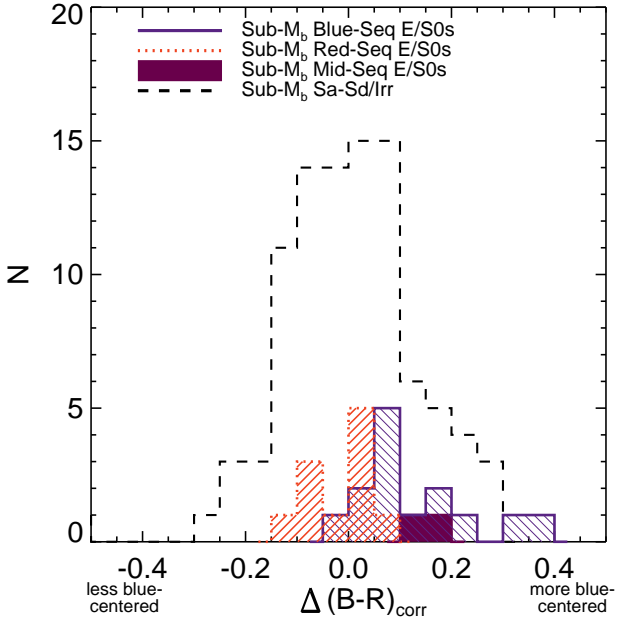}
\caption{Distribution of ${\Delta(B-R)_{\rm corr}}$ for sub-$M_b$ NFGS
  galaxies, where ${\Delta(B-R)_{\rm corr}}$ is the outer disk color
  (between 50--75\% light radii) minus the central color (within 50\%
  light radii), corrected for the typical color gradient of a galaxy
  at that galaxy's blue luminosity (see
  \citealt{kannappan04a}). \label{fig.dbrcorrhist}}
\end{figure}

In \S \ref{section.simulations} above, we presented the possibility of
minor mergers and/or interactions acting as a trigger which induces
gas inflow and star formation. Here we examine observational evidence
that inflow events, most likely triggered externally, are in fact
occurring in blue-sequence E/S0s.

In addition to having bluer inner and outer disk colors than
red-sequence E/S0s, half of the blue-sequence E/S0s in the NFGS have
centers (within 50\% light radii) that are {\it bluer} than their
outer disks (50--75\% light radii) \citepalias{kannappan09a}. Figure
\ref{fig.dbrcorrhist} plots the distribution of a related quantity,
${\Delta(B-R)_{\rm corr}}$, for sub-$M_b$ NFGS
galaxies. ${\Delta(B-R)_{\rm corr}}$ is the color gradient of a galaxy
``corrected'' for the mean color difference of galaxies of the same
$L_B$ \citep{kannappan04a}, since larger galaxies may have central
colors diluted by preexisting red bars or bulges. Figure
\ref{fig.dbrcorrhist} shows that all but one of the sub-$M_b$
blue-sequence E/S0s and both mid-sequence E/S0s are on the more
blue-centered end of the distribution for spiral/irregular galaxies.

In the broader NFGS, larger values of ${\Delta(B-R)_{\rm corr}}$
reflect central star formation enhancements and correlate strongly
with morphological peculiarities and the presence of nearby companions
\citep{kannappan04a}, suggesting a triggered gas inflow
scenario. \citet{kewley06a} also find evidence for gas inflows in
interacting blue-centered galaxies based on radial trends in gas
metallicity. These results are consistent with preliminary evidence
for a relationship between $\Delta(B-R)_{\rm corr}$ and
molecular-to-atomic gas mass ratios in S0--Sb galaxies, to be reported
in \citetalias{kannappan09b}, which implies that HI can in fact flow
inwards and become ${\rm H_2}$, fueling central star formation.

\begin{figure}
\includegraphics[scale=.7,angle=0]{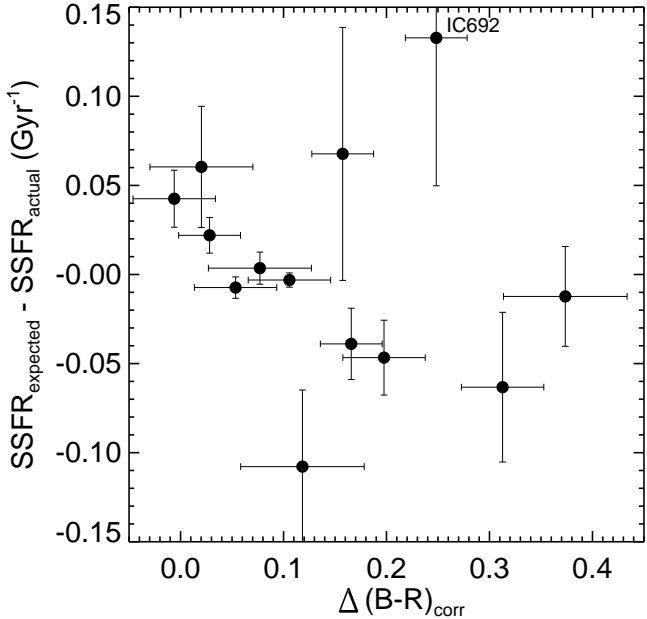}
\caption{Difference between expected specific star formation rate
  (${\rm SSFR_{expected} \equiv } \frac{M_{\rm HI+He}/3\,{\rm Gyr}}{M_*}$)
  and observed specific star formation rate for sub-$M_b$ blue- and
  mid-sequence E/S0s as a function of blue-centeredness. One of our
  blue-sequence E/S0s, UGC~9562, is excluded from this figure because
  it is a polar ring galaxy and not expected to behave normally with
  respect to inflow \citepalias{kannappan09a}. \label{fig.ssfrdbrcorr}}
\end{figure}

If the range of blue-centeredness we see in the sub-$M_b$
blue-sequence E/S0s indicates different burst stages resulting from
episodic gas inflow, then perhaps the specific star formation rate
(SSFR) in these galaxies should scatter about some average SSFR
expected based on the median 3 Gyr gas exhaustion timescale and the
atomic gas mass of each galaxy. We consider the difference between
this expected SSFR (${\rm SSFR_{expected}} = {\rm
  SFR_{expected}}/{M_*} \equiv (M_{\rm HI+He}/3\,{\rm Gyr})/{M_*}\,$)
and the observed SSFR as a function of ${\Delta(B-R)_{\rm corr}}$
(Figure \ref{fig.ssfrdbrcorr}). The error bars for ${\Delta(B-R)_{\rm
    corr}}$ are formal errors, and do not include systematic
uncertainties from ${\Delta(B-R)}$ not being corrected for dust and/or
from delays between the bluest stellar population colors and the peak
of star formation. Such uncertainties may cause some of the
significant scatter seen in this figure. Nonetheless, there is a
suggestive trend with the more blue-centered galaxies having enhanced
SSFRs, and the less blue-centered galaxies having reduced SSFRs.

The Spearman rank correlation coefficient for the data in Figure
\ref{fig.ssfrdbrcorr} is $-0.39$, with the probability of a null
correlation at 19\%. However, if we exclude IC~692 (${\rm
  SSFR_{expected} - SSFR_{observed}}$ = 0.13) from the calculation,
the correlation coefficient becomes $-0.64$, with a probability of a
null correlation decreasing to 3\%. Given the error bars, we infer
that the relationship is likely real. This supports the inflow-driven
burst picture, although follow-up on the apparent outliers would be
informative.

The correlation between blue-centeredness and several properties ---
morphological peculiarities, the presence of companions, increasing
molecular-to-atomic gas mass ratios, and enhanced SSFRs --- suggests
that the inflow of gas in sub-$M_b$ blue-sequence E/S0s may be
episodic and triggered externally. Episodic inflows of gas in
sub-$M_b$ blue-sequence E/S0s would imply that these galaxies are in
various stages of bursty star formation, which is in agreement with
the large ranges of prospective stellar mass growth (\S
\ref{section.massgrowth}) and gas exhaustion times (\S
\ref{section.timescale}) we see for this population.

This result also suggests that our focus on HI rather than molecular
gas (to be discussed in Wei et al., in prep.) does not render our
estimates of stellar mass growth and gas exhaustion time in \S
\ref{section.massgrowth} and \ref{section.timescale} highly
inaccurate. Although stars typically form from molecular gas, the
estimates of inflow and minor merger/interaction timescales suggest
that the transition from HI to ${\rm H_2}$ may occur quickly enough
that the potential for morphological transformation can be realized
even after material already in the molecular phase has been consumed
or otherwise dispersed. These inflow events may also play an important
role in bridging between the sub-$M_b$ red- and blue-sequence E/S0
populations, triggering star formation in sub-$M_b$ red-sequence E/S0s
such as the gas-rich UGC~5923 (discussed in \S \ref{section.outliers})
and moving E/S0s onto the blue sequence.

Observations of the SAURON E/S0 sample \citep{dezeeuw02,emsellem04}
have found that many intermediate-mass early-type galaxies seem to
contain a disk-like stellar component (e.g.,
\citealt{emsellem07,krajnovic08}) --- possibly the remnant of a burst
of star formation induced by the inflow of gas as discussed in this
section. It is possible that some of our sub-$M_b$ blue-sequence E/S0s
are in an active, post gas-infall star-forming phase but will
eventually become passive early-type galaxies like those observed by
\citet{krajnovic08}.

Some of the differences between the SAURON E/S0s and our blue-sequence
E/S0s, however, likely arise from the differences between the two
samples. While there are some E/S0s below the bimodality mass ${M_b
  \sim 3 \times 10^{10}\,M_{\odot}}$ in the SAURON sample, these are
in the minority; in contrast, the disk-building sub-population of
blue-sequence E/S0s becomes numerically important only below this
mass, and more notably below the threshold mass at $5 \times
10^{9}\,M_{\odot}$ (\citetalias{kannappan09a}). It is also possible
that the accretion of fresh gas from external sources, as discussed in
\S\ref{section.simulations}, will reignite star formation in some
red-sequence E/S0s (in both samples) and continue building disks
(\citetalias{kannappan09a}; \citealt{cortese09}). The high frequency
of blue-sequence E/S0s below the threshold mass ($\sim$5\% of the
general galaxy population and $\sim$25\% of E/S0s,
\citetalias{kannappan09a}) would be difficult to explain by quenching
mergers alone, but could be explained if these systems' star-forming
phase is either long or episodic.

\section{Summary}\label{section.discussion}
\citet*{kannappan09a} recently identified a population of E/S0s that
reside on the blue sequence in color vs. stellar mass space, where
spiral galaxies typically reside. Blue-sequence E/S0s increase in
numbers below the bimodality mass ($M_b \sim
3\times10^{10}\,M_{\odot}$) and especially below the gas richness
threshold ($\sim$5$\times10^{9}\,M_{\odot}$). These sub-$M_b$
blue-sequence E/S0s are characterized by fairly regular morphology,
and many appear to be rebuilding disks
\citepalias{kannappan09a}. Blue-sequence E/S0s also fall between
red-sequence E/S0s and spirals in the $M_*$-radius and $M_*$-$\sigma$
relations, suggesting that they may be a transitional population
\citepalias{kannappan09a}.

In this paper, we examined the atomic gas content of blue-sequence
E/S0s below $\sim$$M_b$ to determine whether they have large enough
gas reservoirs to transform to later-type morphology. In a
representative sample drawn from the Nearby Field Galaxy Survey, we
find that blue-sequence E/S0s have substantial atomic gas masses in
the range of $10^{7}$--$10^{10}\, {M_{\odot}}$, comparable to the gas
masses of spiral and irregular galaxies in the same stellar mass
range. Blue- and red-sequence E/S0s have distinctly different atomic
gas-to-stellar mass ratios, with most blue-sequence E/S0s in the range
of 0.1--1.0 and most red-sequence E/S0s $<$0.1. This suggests
significantly greater potential for morphological transformation in
blue-sequence E/S0s than red-sequence E/S0s.

Combining these atomic gas masses with current rates of star
formation, we find that many of the sub-$M_b$ blue-sequence E/S0s can
form new stars in the range of 10--60$\%$ of their current stellar
masses within 3 Gyr, in both constant (i.e., allowing gas infall) and
exponentially declining (i.e., closed box) star formation scenarios,
provided that the atomic gas reservoir is available for star
formation. In the constant star formation scenario, we find that about
half of the sub-$M_b$ blue-sequence E/S0 systems will exhaust their
gas reservoirs in $\lesssim$3 Gyr if no fresh gas infall is permitted.

Because of the lack of spatial resolution in our HI data, we cannot
say for certain that the gas is readily available for star
formation. We find evidence, however, which indicate that fresh gas
may be brought inwards and made available for star formation on
timescales shorter than the gas exhaustion timescale. We estimate the
dynamical timescale for infall of extra-planar gas to be on average
$<$1 Gyr, shorter than the gas exhaustion time for most of our
galaxies. The frequency of inflow for gas trapped in diffuse disks may
be dominated by the rate of minor mergers/interactions, which
simulations find to occur every 1.4 Gyr or less for progenitor mass
ratios of 1:100 or smaller \citep{fakhouri08}. As evidence of such
events, we find that sub-$M_b$ blue-sequence E/S0s are more often
blue-centered than the general galaxy population, where
blue-centeredness is measured relative to the typical color gradient
of galaxies at a given luminosity. \citet{kannappan04a} find
blue-centeredness to correlate with morphological peculiarities and
companions, which supports the externally triggered gas inflow
scenario. For blue-sequence E/S0s, we find a relationship between
blue-centeredness and variations in specific star formation rates
relative to typical reference values, suggesting that such inflows may
be episodic and trigger bursts of star formation. In summary, this
work clearly confirms that blue-sequence E/S0s have both the gas
reservoirs and the potential for sustained star formation necessary
for significant disk growth, consistent with evolution toward
later-type morphology if the spatial distribution of the gas is
extended. While our sample of 27 blue- and red-sequence E/S0s is
sufficient for the analysis presented in this paper, it is important
to extend our work to a larger sample of sub-$M_b$ E/S0s for more
robust statistics. The multi-wavelength RESOLVE Survey underway
\citep{kannappan08conf} would be an ideal data set for such a study.

Active follow-up includes obtaining IRAM CO(1--0) and CO(2--1) spectra
to quantify the molecular gas content in these galaxies, which may be
a large fraction of the gas mass content and further extend the
potential for morphological transformation in blue-sequence E/S0s. We
are also obtaining VLA HI and CARMA CO(1--0) maps to resolve the
distribution of the atomic and molecular gas in blue-sequence
E/S0s. HI maps will allow us to look for extended gas disks, small
companions, and/or signs of interactions. CO(1--0) maps may reveal
inner disky ``pseudobulges'' growing in tandem with extended disks, as
would be expected during the formation of late type galaxies.


\acknowledgements 

We thank the anonymous referee for his/her helpful comments.  We are
grateful to A. Bolatto, E. Gawiser, S. Khochfar, D. Mar, S. McGaugh,
J. Rose, E. Shaya, and P. Teuben for insightful conversations. We
would like to thank the GBT operators and the Green Bank staff for
support during this program (07A-072, 07C-148). The National Radio
Astronomy Observatory is a facility of the National Science Foundation
operated under cooperative agreement by Associated Universities,
Inc. This material is based upon work supported by the National
Science Foundation under Grant No. AST-0838178. This research has made
use of the NASA/IPAC Extragalactic Database (NED) which is operated by
the Jet Propulsion Laboratory, California Institute of Technology,
under contract with the National Aeronautics and Space
Administration. We acknowledge the usage of the HyperLeda database
(http://leda.univ-lyon1.fr) and the Cornell Digital HI Archive
(http://arecibo.tc.cornell.edu/hiarchive). The Parkes telescope is
part of the Australia Telescope which is funded by the Commonwealth of
Australia for operation as a National Facility managed by CSIRO.


\appendix
\section{Appendix A}\label{section.appendix}

\subsection{GBT Flux Measurements and Comparison with Literature Fluxes}
\begin{figure}
\includegraphics[scale=.7,angle=0]{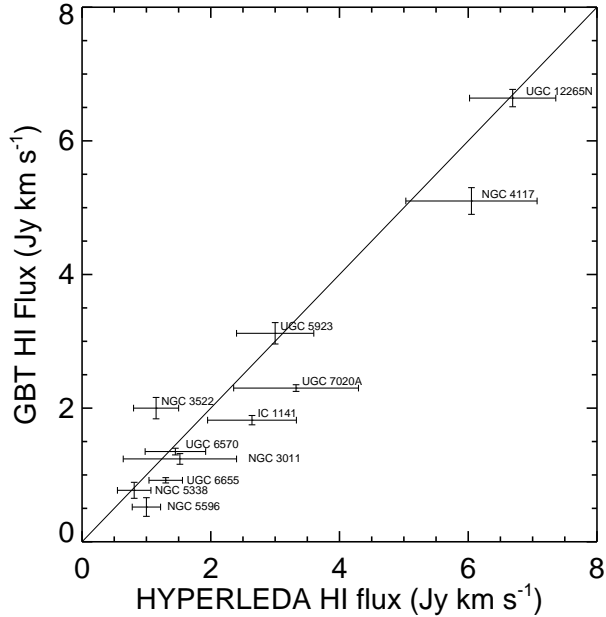}
\caption{Comparison of HI fluxes between our new GBT observations and
  HyperLeda homogenized data.\label{fig.fluxdiff}}
\end{figure}

Figure \ref{fig.fluxdiff} plots our new GBT HI fluxes against
literature fluxes compiled in HyperLeda \citep{paturel03b}, showing
good agreement between the two. Notes on individual galaxies are as
follows:

For many galaxies (UGC~6655, UGC~7020A, NGC~3011, IC~1141, UGC~6570)
the GBT HI profiles have much stronger S/N than those from the
literature, hence any differences in flux measurements between GBT and
literature are likely to be attributed to the noisy profiles of the
literature data. This can be seen in the error bars in Figure
\ref{fig.fluxdiff}.

NGC~5596: the literature flux is very noisy and has a velocity width
of $>500\, {\rm km\,s^{-1}}$, so it makes sense that our flux
measurement is smaller considering we measure a smaller velocity width.

NGC~4117: the literature data are also of poorer quality and suffer
from confusion with a nearby companion NGC~4118 (1.5$\arcmin$ and 643
${\rm km\,s^{-1}}$). We separate out the companion by using the width
of the ionized gas from 850--1050 ${\rm km\,s^{-1}}$, which also agrees
with our preliminary CO data for this galaxy.

UGC~12265N: one member of a galaxy pair; the HI is completely blended
in with its smaller companion, UGC~12265S, only $1\arcmin$ and
$\sim$70 ${\rm km\,s^{-1}}$ away (see Figure
\ref{fig.gbtspectra_blue}). Preliminary VLA HI data suggest that
UGC~12265N contains about 1/4 of total HI within the GBT beam, so we
use this fraction of the total HI flux for our analysis.

NGC~3522: HI spectrum from \citet{lake84} seem to have comparable S/N
to our GBT spectrum, due to our rather short integration time. The
$\sim$3$\arcmin$ beam of the Arecibo Telescope used by \citet{lake84},
however, could be missing some extended HI flux, which would explain
the higher flux measurement on our part.

IC~1639, IC~195: both have known companions within the beam of the
GBT. Because these galaxies do not have ionized-gas rotation
information, we measure the HI flux within the range of stellar
rotation (5344--5444 ${\rm km\,s^{-1}}$ and 3498--3798 ${\rm
  km\,s^{-1}}$, respectively).

NGC~1029: does not have ionized-gas or stellar rotation data, so we
exclude the companion by measuring the HI flux within a width given by
Arecibo HI observations of this galaxy. The smaller 3$\arcmin$ beam of
the Arecibo Telescope does not detect the companion and finds a
${W_{20}}$ of 353 ${\rm km\,s^{-1}}$ \citep{springob05}.

The following galaxies are classified as undetected, though HI
emission was detected in velocity ranges corresponding to nearby
galaxies. For these three galaxies, we give both the HI upper limit
for the target galaxy as well as the HI flux from the nearby galaxy.

NGC~4308: The measured HI velocity suggests that the flux belongs to a
nearby companion, UGC~7438 ($\sim$5$\arcmin$, 699 ${\rm km\,s^{-1}}$),
though this is not certain. Assigning the gas to NGC~4308 does not
change any results since the measured atomic gas mass is still
extremely low, $< 1\%$ of the stellar mass of the galaxy. 

IC~1144: The measured HI velocity suggests that the flux belongs to a
nearby companion, SDSS J155124.19+432506.8 ($\sim$0.5$\arcmin$, 12225
${\rm km\,s^{-1}}$).

UGC~12835: Based on the measured HI velocity, the HI flux clearly
belongs to a companion in the beam, and the target galaxy is
undetected.

\subsection{GBT HI Profiles of Sub-$M_b$ E/S0s}
We examine and discuss the HI profiles of red- and blue-sequence
E/S0s, grouping them by different types of velocity profiles while
noting the sequence they belong to (B, mid-sequence, and R).

{\it Galaxies with flat velocity profiles:} NGC~3011(B), NGC~1552(R),
NGC~3522(R), and NGC~3499(R) all have HI profiles that are reminiscent
of a gas disk extending either slightly into or well beyond the flat
part of the rotation curve. These profiles are not as sharply peaked
as the typical double-horned profile, suggesting that the gas disk
does not extend too far into the flat part of the rotation curve. None
of these galaxies have large companions within the beam of the GBT,
though all of these profiles seem to have slight asymmetries,
suggesting there may be distortions in the gas disk.

{\it Galaxies with rising velocity profiles:} UGC~6003(B),
UGC~6655(B), UGC~6570(Mid), UGC~7020A(Mid), and NGC~5338(R) have
sharply peaked HI profiles, which are indicative of gas disks that are
still in the rising part of the rotation curve. Most of these are
relatively symmetric and do not have any known large companions within
the beam of the GBT.

{\it Galaxies in between the first two cases:} UGC~6805(B),
IC~1141(B), and UGC~5923(R) have HI profiles that seem to be a
combination of the previous two cases --- their HI profiles all have a
single peak like the rising rotation curve case, but there is a slight
flattening suggesting that the gas is reaching the flat part of the
rotation curve. UGC~6805 seems to have a gas disk with rotation in the
rising part of the curve; although the center seems slightly
flattened, it is hard to tell given the noise in the HI
profile. IC~1141 and UGC~5923, on the other hand, seem to have centers
that are much more flattened, suggesting that the gas disk may extend
a little further into the flat part of the rotation curve. These
galaxies also do not have any known large companions.

{\it Galaxies with known companions: }NGC~4117(R), UGC~12265N(B),
IC~1639(B), and IC~195(B) all have optically confirmed
companions. NGC~4117 has an asymmetric double horn profile between 800
and 1100 ${\rm km\,s^{-1}}$, and a companion (NGC~4118) $1.5\arcmin$
and 643 ${\rm km\,s^{-1}}$ away. UGC~12265N is actually a galaxy pair
with UGC~12265S, and its HI profile is completely blended in since the
two are less than 100 ${\rm km\,s^{-1}}$ apart in velocity space. The
larger peak in the HI profile for IC~1639 is actually its larger
companion, IC~1640; IC~1639 is the smaller peak at 2581 ${\rm
  km\,s^{-1}}$. IC~195 has two companions: IC~196, which is
$2.2\arcmin$ away, but only about 15 ${\rm km\,s^{-1}}$ apart in
velocity space (so it is completely blended in), and Arp 290, which
probably contributes to the HI profile at 3509 ${\rm km\,s^{-1}}$,
located $1.1\arcmin$ away.

{\it Non-detections:} UGC~130(R), NGC~516(R), NGC~3179(R),
NGC~3605(R), NGC~4308(R), UGC~8876(R), IC~1144(B), and UGC~12835(R).

{\it Miscellaneous:} NGC~5596(R) and NGC 1298 (R) are both detections,
but the S/N of their spectra are too poor to allow us to categorize
their HI profiles.



\clearpage
\LongTables

\begin{deluxetable}{lllcccccccccc}
\tabletypesize{\scriptsize}
\tablewidth{0pt}
\tablecaption{HI Data for the NFGS\label{table.nfgshi}}
\tablehead{
\colhead{ID}               
& \colhead{Galaxy Name}      
& \colhead{UGC}      
& \colhead{Morph.}
& \colhead{Sequence}
& \colhead{$V_{\odot}$($\sigma_{V_{\odot}}$)}  
& \colhead{$V_{\rm M}^{{\rm sin} i}$($\sigma_{V_{\rm M}^{{\rm sin} i}}$)}  
& \colhead{$f_{\rm HI}(\sigma_{f_{\rm HI}})$}               
& \colhead{$M_{\rm HI}$}
& \colhead{sub-$M_b$ E/S0}       
& \colhead{Reference} 
\\
\colhead{}                 
& \colhead{}                 
& \colhead{} 
& \colhead{}          
& \colhead{} 
& \colhead{(km\,${\rm s^{-1}}$)}        
& \colhead{(km\,${\rm s^{-1}}$)}           
& \colhead{(Jy\,km\,${\rm s^{-1}}$)}
& \colhead{($M_{\odot}$)} 
& \colhead{Sample}
& \colhead{} 
\\
\colhead{(1)}               
& \colhead{(2)}      
& \colhead{(3)}      
& \colhead{(4)}  
& \colhead{(5)}  
& \colhead{(6)}  
& \colhead{(7)}              
& \colhead{(8)}    
& \colhead{(9)}
& \colhead{(10)}    
& \colhead{(11)} 
}

\startdata

1    & A00113p3037  & 130   & cE   &  R &	   &	      & $<$0.12      & $<$1.36$    \times 10^8$     &	       &  GBT   \\
2    & A00289p0556  & 313   & Sd   &  B & 2085(7)  & 48(2)    & 0.78(0.13)   & 1.65$\pm0.27\times 10^8$     &	       &  LE    \\
3    & NGC193	    & 408   & L-   &  R & 4220?    &	      & $<$2.0       & $<$1.80$    \times 10^9$     &	       &  HR   \\
4    & A00389m0159  & 439   & Sa   &  B & 5299(7)  & 55(3)    & 4.94(1.28)   & 6.54$\pm1.69\times 10^9$     &	       &  LE    \\
5    & A00442p3224  & 484   & Sb   &  B & 4857(8)  & 197(6)   & 13.12(4.83)  & 1.56$\pm0.57\times 10^{10}$  &	       &  LE    \\
6    & A00510p1225  & 545   & Sc   &    & 18339(8) & 175(9)   & 1.03(0.33)   & 1.66$\pm0.53\times 10^{10}$  &	       &  LE    \\
7    & NGC315	    & 597   & E    &  R & 4921     &	      & $<$1.33      & $<$1.63$    \times 10^9$     &	       &  HR   \\
8    & A00570p1504  & 615   & Sab  &  B & 5507(3)  & 164(12)  & 1.72(0.21)   & 2.53$\pm0.31\times 10^9$     &	       &  CEGG   \\
9    & A01047p1625  & 685   & Im   &    & 156(4)   & 32(1)    & 11.32(2.04)  & 5.07$\pm0.91\times 10^7$     &	       &  LE    \\
10   & NGC382	    & 688   & E    &  R &	   &	      & 	     &  			    &	       &  LE    \\
11   & IC1639	    & 750   & cE   &  B & 5381(12) & 22(10)   & 0.70(0.08)   & 9.55$\pm1.09\times 10^8$     & $\times$ &  GBT   \\
12   & A01123m0046  & 793   & Sc   &  B &	   &	      & $<$1.98	     & $<$1.01$	   \times 10^{10}$  &	       &  HI  \\
13   & A01187m0048  & 892   & Sa   &  B & 5228(5)  & 32(2)    & 2.37(0.33)   & 3.04$\pm0.42\times 10^9$     &	       &  CEGG   \\
14   & NGC516	    & 946   & L    &  R &	   &	      & $<$0.09      & $<$2.59$    \times 10^7$     & $\times$ &  GBT   \\
15   & A01300p1804  & 1104  & Im   &  B & 684(4)   & 44(2)    & 7.62(0.97)   & 2.35$\pm0.30\times 10^8$     &	       &  LE    \\
16   & A01344p2838  & 1154  & Sbc  &  B & 7758(7)  & 124(5)   & 5.03(2.38)   & 1.47$\pm0.70\times 10^{10}$  &	       &  LE    \\
17   & A01346p0438  & 1155  & Sbc  &  B & 3181(8)  & 85(4)    & 2.50(0.42)   & 1.18$\pm0.20\times 10^9$     &	       &  LE    \\
18   & A01374p1539B & 1176  & Im   &    & 632(3)   & 16(0.4)  & 23.23(2.24)  & 5.98$\pm0.58\times 10^8$     &	       &  LE    \\
19   & NGC695	    & 1315  & Sc   &  B & 9729(6)  & 148(5)   & 3.93(0.84)   & 1.84$\pm0.39\times 10^{10}$  &	       &  LE    \\
20   & NGC784	    & 1501  & Sm   &    & 198(4)   & 41(1)    & 45.50(4.39)  & 2.87$\pm0.28\times 10^8$     &	       &  LE    \\
21   & A02008p2350  & 1551  & Sdm  &  B & 2671(5)  & 47(2)    & 11.53(1.11)  & 4.15$\pm0.40\times 10^9$     &	       &  LE    \\
22   & IC195	    & 1555  & S0/a &  B & 3648(2)  & 104(7)   & 6.85(0.16)   & 4.39$\pm0.10\times 10^9$     & $\times$ &  GBT   \\
23   & IC197	    & 1564  & Sbc  &  B & 6316(6)  & 122(9)   & 8.26(0.86)   & 1.55$\pm0.16\times 10^{10}$  &	       &  CEGG   \\
24   & IC1776	    & 1579  & Sc   &  B & 3410(6)  & 48(2)    & 12.19(3.16)  & 6.64$\pm1.72\times 10^9$     &	       &  LE    \\
25   & A02056p1444  & 1630  & Sb   &  B & 4437(3)  & 133(10)  & 6.39(0.67)   & 5.94$\pm0.62\times 10^9$     &	       &  CEGG   \\
26   & NGC825	    & 1636  & Sa   &  R & 3397(5)  & 217(10)  & 5.89(1.52)   & 3.20$\pm0.83\times 10^9$     &	       &  LE    \\
27   & NGC927	    & 1908  & Sc   &  B & 8266(10) & 94(4)    & 4.43(1.15)   & 1.43$\pm0.37\times 10^{10}$  &	       &  LE    \\
28   & A02257m0134  & 1945  & Sdm  &  B & 1768(8)  & 49(3)    & 2.30(0.2)    & 3.25$\pm0.28\times 10^8$     &	       &  LE    \\
29   & NGC984	    & 2059  & Sa   &  R & 4355(7)  & 278(18)  & 10.62(4.59)  & 1.00$\pm0.43\times 10^{10}$  &	       &  LE    \\
30   & NGC1029      & 2149  & S0/a &  R & 3620(28) & 127(9)   & 8.32(0.15)   & 5.19$\pm0.09\times 10^9$     &	       &  GBT   \\
31   & A02464p1807  & 2296  & cE   &    &	   &	      & $<$6.9       & 3.39$	   \times 10^{10}$  &	       &  HR   \\
32   & A02493m0122  & 2345  & Sm   &  B & 1506(5)  & 40(2)    & 19.68(2.5)   & 2.03$\pm0.26\times 10^9$     &	       &  LE    \\
33   & NGC1298      & 2683  & ?E   &  R & 6452(3)  & 36(2)    & 0.48(0.09)   & 9.43$\pm1.77\times 10^8$     &	       &  GBT   \\
34   & A03202m0205  & 2704  & Sa   &  B &	   &	      & $<$1.92	     & $<$6.03$	   \times 10^9$     &	       &  HI  \\
35   & NGC1552      & 3015  & L    &  R & 4784(8)  & 129(9)   & 0.7(0.09)    & 7.76$\pm1   \times 10^8$     &	       &  GBT   \\
36   & NGC2692      & 4675  & Sa   &  R & 4026(12) & 75(4)    & 3.03(0.42)   & 2.37$\pm0.33\times 10^9$     &	       &  LE    \\
37   & A08567p5242  & 4713  & Sb   &  B & 9036(7)  & 249(9)   & 3.09(0.27)   & 1.27$\pm0.11\times 10^{10}$  &	       &  LE    \\
38   & A09045p3328  & 4787  & Sdm  &  B & 552(8)   & 53(1)    & 6.00(1.02)   & 1.07$\pm0.18\times 10^8$     &	       &  LE    \\
39   & NGC2780      & 4843  & Sab  &  B & 1962(7)  & 130(10)  & 0.25(0.06)   & 5.83$\pm1.40\times 10^7$     &	       &  LE    \\
40   & A09125p5303  & 4879  & Im   &  R &	   &	      & 	     &  			    &	       &  LE    \\
41   & NGC2799      & 4909  & Sm   &  B & 1755(25) & 146(13)  & 7.55(2.5)    & 1.70$\pm0.56\times 10^9$     &	       &  LE    \\
42   & NGC2824      & 4933  & L    &  R &	   &	      & 	     &  			    &	       &  LE    \\
43   & NGC2844      & 4971  & Sa   &  B & 1489(6)  & 145(5)   & 6.46(0.82)   & 9.75$\pm1.24\times 10^8$     &	       &  LE    \\
44   & NGC3011      & 5259  & S0/a &  B & 1543(2)  & 76(5)    & 1.24(0.08)   & 1.93$\pm0.12\times 10^8$     & $\times$ &  GBT   \\
45   & NGC3009      & 5264  & Sc   &  B & 4564(12) & 194(13)  & 2.95(0.34)   & 3.42$\pm0.39\times 10^9$     &	       &  LE    \\
46   & IC2520	    & 5335  & Pec  &  B & 1238(8)  & 85(8)    & 6.00(1.69)   & 6.31$\pm1.78\times 10^8$     &	       &  LE    \\
47   & A09557p4758  & 5354  & Sm   &  B & 1171(4)  & 76(3)    & 17.14(1.48)  & 1.85$\pm0.16\times 10^9$     &	       &  LE    \\
48   & NGC3075      & 5360  & Sc   &  B & 3582(8)  & 124(4)   & 15.07(7.12)  & 1.00$\pm0.47\times 10^{10}$  &	       &  LE    \\
49   & A09579p0439  & 5378  & Sb   &  B & 4161(10) & 117(4)   & 6.05(2.86)   & 5.31$\pm2.51\times 10^9$     &	       &  LE    \\
50   & NGC3104      & 5414  & Im   &  B & 603(3)   & 41(2)    & 17.78(1.36)  & 4.33$\pm0.33\times 10^8$     &	       &  LE    \\
51   & A10042p4716  & 5451  & Im   &  R & 629(9)   & 52(2)    & 2.84(0.64)   & 7.17$\pm1.62\times 10^7$     &	       &  LE    \\
52   & NGC3165      & 5512  & Im   &  R & 1333(4)  & 61(2)    & 3.45(0.78)   & 3.63$\pm0.82\times 10^8$     &	       &  LE    \\
53   & A10114p0716  & 5522  & Sc   &  B & 1220(4)  & 90(2)    & 27.92(2.98)  & 2.62$\pm0.28\times 10^9$     &	       &  LE    \\
54   & NGC3179      & 5555  & L    &  R &	   &	      & $<$0.19      & $<$5.10$    \times 10^8$     &	       &  GBT   \\
55   & A10171p3853  & 5577  & Sm   &  B & 2028(9)  & 45(2)    & 3.30(0.93)   & 8.61$\pm2.43\times 10^8$     &	       &  LE    \\
56   & NGC3213      & 5590  & Sbc  &  B & 1346(5)  & 67(4)    & 1.47(0.35)   & 2.01$\pm0.48\times 10^8$     &	       &  LE    \\
57   & NGC3264      & 5719  & Sdm  &  B & 940(7)   & 60(2)    & 14.93(1.44)  & 1.16$\pm0.11\times 10^9$     &	       &  LE    \\
58   & NGC3279      & 5741  & Sc   &  B & 1393(6)  & 156(4)   & 3.55(0.96)   & 4.81$\pm1.3 \times 10^8$     &	       &  LE    \\
59   & A10321p4649  & 5744  & Sc   &  B &	   &	      & 	     &  			    &	       &  LE    \\
60   & A10337p1358  & 5760  & Scd  &  B & 3010(6)  & 139(5)   & 3.78(0.33)   & 1.86$\pm0.16\times 10^9$     &	       &  LE    \\
61   & IC2591	    & 5763  & Sbc  &  B & 6797(10) & 160(5)   & 8.75(4.13)   & 2.04$\pm0.96\times 10^{10}$  &	       &  LE    \\
62   & A10365p4812  & 5791  & Sc   &  B & 857(5)   & 67(3)    & 3.72(1.05)   & 2.34$\pm0.66\times 10^8$     &	       &  LE    \\
63   & A10368p4811  & 5798  & Sc   &  B & 1519(6)  & 72(5)    & 5.37(1.52)   & 9.36$\pm2.65\times 10^8$     &	       &  LE    \\
64   & NGC3326      & 5799  & Sb   &  B & 8153(8)  & 100(6)   & 2.77(0.88)   & 8.97$\pm2.85\times 10^9$     &	       &  LE    \\
65   & A10389p3859  & 5819  & Sbc  &  B &	   &	      & 	     &  			    &	       &  LE    \\
66   & A10431p3514  & 5870  & Sa   &  B & 1992(6)  & 100(7)   & 4.59(1.3)    & 1.24$\pm0.35\times 10^9$     &	       &  LE    \\
67   & A10448p0731  & 5892  & Sb   &  B &	   &	      & $<$1.53	     & $<$4.97$    \times 10^9$     &	       &  HI  \\
68   & A10465p0711  & 5923  & S0/a &  R & 713(4)   & 65(5)    & 3.12(0.16)   & 4.71$\pm0.24\times 10^7$     & $\times$ &  GBT   \\
69   & A10504p0454  & 6003  & S0/a &  B & 5819(5)  & 67(5)    & 1.47(0.09)   & 2.46$\pm0.15\times 10^9$     & $\times$ &  GBT   \\
70   & NGC3454      & 6026  & Sc   &  B & 1109(5)  & 91(3)    & 5.62(1.26)   & 5.62$\pm1.26\times 10^8$     &	       &  LE    \\
71   & A10592p1652  & 6104  & Sbc  &  B & 2947(5)  & 110(2)   & 7.41(1.26)   & 3.59$\pm0.61\times 10^9$     &	       &  LE    \\
72   & NGC3499      & 6115  & S0/a &  R & 1495(12) & 119(9)   & 0.39(0.06)   & 7.15$\pm1.10\times 10^7$     & $\times$ &  GBT   \\
73   & NGC3510      & 6126  & Sd   &  B & 704(3)   & 79(2)    & 33.57(2.9)   & 7.26$\pm0.63\times 10^8$     &	       &  LE    \\
74   & Mrk421	    & 6132W & Pec  &    &	   &	      & 	     &  			    &	       &  LE    \\
75   & NGC3522      & 6159  & L-   &  R & 1221(8)  & 112(8)   & 2.00(0.16)   & 2.47$\pm0.20\times 10^8$     & $\times$ &  GBT   \\
76   & A11040p5130  & 6162  & Sc   &  B & 2208(5)  & 95(4)    & 12.65(1.74)  & 4.04$\pm0.56\times 10^9$     &	       &  LE    \\
77   & IC673	    & 6200  & Sa   &  B & 3859(8)  & 151(5)   & 17.14(8.09)  & 1.31$\pm0.62\times 10^{10}$  &	       &  LE    \\
78   & A11068p4705  & 6201  & L-   &  R &	   &	      & 	     &  			    &	       &  LE    \\
79   & A11072p1302  & 6206  & Sc   &  B &	   &	      & 	     &  			    &	       &  LE    \\
80   & NGC3605      & 6295  & L-   &  R &	   &	      & $<$0.11      & $<$1.61$    \times 10^6$     & $\times$ &  GBT   \\
81   & A11142p1804  & 6296  & Sc   &  R & 976(6)   & 82(3)    & 2.10(0.72)   & 8.28$\pm2.84\times 10^7$     &	       &  LE    \\
82   & NGC3633      & 6351  & Sa   &  B & 2599(6)  & 150(5)   & 2.90(0.46)   & 1.07$\pm0.17\times 10^9$     &	       &  LE   \\
83   & IC692	    & 6438  & E    &  B & 1163(8)  & 35(3)    & 2.50(0.71)   & 2.71$\pm0.77\times 10^8$     & $\times$ &  LE    \\
84   & A11238p5401  & 6446  & Sd   &    & 645(4)   & 57(2)    & 28.71(3.65)  & 1.13$\pm0.14\times 10^9$     &	       &  LE    \\
85   & A11310p3254  & 6545  & Sb   &  B & 2630(6)  & 66(7)    & 1.00(0.28)   & 4.20$\pm1.18\times 10^8$     &	       &  LE    \\
86   & IC708	    & 6549  & ?E   &  R &	   &	      & 	     &  			    &	       &  LE    \\
87   & A11332p3536  & 6570  & S0/a & Mid& 1628(2)  & 49(3)    & 1.35(0.05)   & 2.60$\pm0.1 \times 10^8$     & $\times$ &  GBT   \\
88   & A11336p5829  & 6575  & Sc   &  B & 1216(5)  & 95(3)    & 13.12(1.27)  & 1.68$\pm0.16\times 10^9$     &	       &  LE    \\
89   & NGC3795A     & 6616  & Scd  &  B & 1148(5)  & 45(2)    & 11.86(1.63)  & 1.39$\pm0.19\times 10^9$     &	       &  LE    \\
90   & A11372p2012  & 6625  & Sc   &  B &	   &	      & 	     &  			    &	       &  LE    \\
91   & NGC3795      & 6629  & Sc   &  B & 1212(6)  & 103(4)   & 7.21(0.62)   & 7.72$\pm0.66\times 10^8$     &	       &  LE    \\
92   & A11378p2840  & 6637  & L-   &  B & 1836(6)  & 61(5)    & 1.84(0.87)   & 4.31$\pm2.04\times 10^8$     & $\times$ &  LE    \\
93   & A11392p1615  & 6655  & L    &  B & 744(2)   & 26(2)    & 0.92(0.04)   & 1.70$\pm0.07\times 10^7$     & $\times$ &  GBT   \\
94   & NGC3846      & 6706  & Sm   &  B & 1451(8)  & 66(3)    & 6.34(0.87)   & 1.00$\pm0.14\times 10^9$     &	       &  LE    \\
95   & NGC3850      & 6733  & Sc   &    & 1149(8)  & 69(3)    & 7.91(1.34)   & 9.43$\pm1.60\times 10^8$     &	       &  LE    \\
96   & A11476p4220  & 6805  & L    &  B & 1132(6)  & 49(3)    & 0.39(0.05)   & 3.78$\pm0.48\times 10^7$     & $\times$ &  GBT   \\
97   & NGC3913      & 6813  & Sd   &    & 954(3)   & 19(1)    & 10.52(1.12)  & 9.06$\pm0.96\times 10^8$     &	       &  LE    \\
98   & IC746	    & 6898  & Sb   &  B & 5028(3)  & 123(4)   & 6.95(1.1)    & 9.31$\pm1.47\times 10^9$     &	       &  LE    \\
99   & A11531p0132  & 6903  & Sc   &    & 1892(6)  & 78(3)    & 15.35(2.44)  & 3.54$\pm0.56\times 10^9$     &	       &  LE    \\
100  & NGC3978      & 6910  & Sbc  &  B & 9962(8)  & 77(3)    & 3.33(0.71)   & 1.64$\pm0.35\times 10^{10}$  &	       &  LE    \\
101  & A11547p4933  & 6930  & Sc   &    & 778(5)   & 50(2)    & 27.41(4.06   & 1.53$\pm0.23\times 10^9$     &	       &  LE    \\
102  & A11547p5813  & 6931  & Sm   &  B & 1195(8)  & 49(2)    & 4.63(0.89)   & 5.62$\pm1.08\times 10^8$     &	       &  LE    \\
103  & NGC4034      & 7006  & Sc   &  B & 2367(8)  & 91(6)    & 3.21(0.28)   & 1.19$\pm0.10\times 10^9$     &	       &  LE    \\
104  & A11592p6237  & 7009  & Im   &  B & 1120(12) & 74(6)    & 6.00(1.02)   & 6.79$\pm1.15\times 10^8$     &	       &  LE    \\
105  & A12001p6439  & 7020A & L    & Mid& 1515(1)  & 39(3)    & 2.30(0.05)   & 3.86$\pm0.08\times 10^8$     & $\times$ &  GBT   \\
106  & NGC4117      & 7112  & L    &  R & 934(2)   & 122(9)   & 5.10(0.20)   & 4.36$\pm0.17\times 10^8$     & $\times$ &  GBT   \\
107  & NGC4120      & 7121  & Sc   &  B & 2246(7)  & 92(4)    & 10.42(0.9)   & 3.53$\pm0.30\times 10^9$     &	       &  LE    \\
108  & A12064p4201  & 7129  & Sab  &  R & 947(9)   & 60(6)    & 1.22(0.25)   & 9.69$\pm1.99\times 10^7$     &	       &  LE    \\
109  & NGC4141      & 7147  & Sc   &  R & 1902(7)  & 70(3)    & 6.89(0.60)   & 1.90$\pm0.17\times 10^9$     &	       &  LE    \\
110  & NGC4159      & 7174  & Sdm  &  B & 1753(12) & 77(5)    & 5.68(2.09)   & 1.29$\pm0.47\times 10^9$     &	       &  LE    \\
111  & NGC4204      & 7261  & Sdm  &    & 858(9)   & 36(1)    & 24.10(4.61)  & 5.80$\pm1.11\times 10^8$     &	       &  LE    \\
112  & NGC4238      & 7308  & Sc   &  B & 2765(6)  & 113(6)   & 5.78(0.50)   & 2.78$\pm0.24\times 10^9$     &	       &  LE    \\
113  & NGC4248      & 7335  & Sdm  &  R & 484(14)  & 29(3)    & 4.63(1.31)   & 8.05$\pm2.28\times 10^7$     &	       &  LE    \\
114  & A12167p4938  & 7358  & Sc   &  B & 3658(6)  & 132(5)   & 5.27(1.01)   & 4.04$\pm0.78\times 10^9$     &	       &  LE    \\
115  & NGC4272      & 7378  & L-   &    &	   &	      & 	     &  			    &	       &  LE    \\
116  & NGC4288      & 7399  & Sm   &  B & 535(4)   & 77(3)    & 26.18(3.6)   & 5.38$\pm0.74\times 10^8$     &	       &  LE    \\
117  & NGC4308      & 7426  & L    &  R &	   &	      & $<$0.06      & $<$1.01$    \times 10^6$     & $\times$ &  GBT   \\
118  & A12195p3222  & 7428  & Sdm  &    & 1138(6)  & 28(1)    & 7.21(2.04)   & 9.43$\pm2.67\times 10^8$     &	       &  LE    \\
119  & A12195p7535  &	    & cE   &    &	   &	      & 	     &  			    &	       &  LE    \\
120  & A12263p4331  & 7608  & Im   &    & 536(5)   & 24(1)    & 20.42(2.39)  & 3.80$\pm0.45\times 10^8$     &	       &  LE    \\
121  & A12295p4007  & 7678  & Sd   &  B & 682(9)   & 24(1)    & 5.03(1.42)   & 1.49$\pm0.42\times 10^8$     &	       &  LE    \\
122  & A12300p4259  & 7690  & Sdm  &  B & 538(11)  & 35(2)    & 14.52(2.46)  & 2.73$\pm0.46\times 10^8$     &	       &  LE    \\
123  & A12304p3754  & 7699  & Sd   &  B & 500(7)   & 81(3)    & 20.04(2.34)  & 2.67$\pm0.31\times 10^8$     &	       &  LE    \\
124  & NGC4509      & 7704  & Sm   &  B & 937(8)   & 38(4)    & 3.89(1.84)   & 2.72$\pm1.28\times 10^8$     &	       &  LE    \\
125  & A12331p7230  & 7761  & Sb   &  B &	   &	      & 	     &  		            &	       &  LE    \\
126  & A12446p5155  & 7950  & Im   &  B & 498(4)   & 35(1)    & 5.89(0.63)   & 1.37$\pm0.15\times 10^8$     &	       &  LE    \\
127  & NGC4758      & 8014  & Sbc  &  B & 1243(6)  & 79(3)    & 8.59(1.46)   & 1.24$\pm0.21\times 10^9$     &	       &  LE    \\
128  & NGC4795      & 8037  & Sa   &  R & 2812(5)  & 155(7)   & 1.70(0.23)   & 7.88$\pm1.07\times 10^8$     &	       &  LE    \\
129  & NGC4807      & 8049  & L+   &  R &	   &	      & 	     &  			    &	       &  LE    \\
130  & NGC4841B     & 8073  & E    &    &	   &	      & 	     &  			    &	       &  LE    \\
131  & NGC4926      & 8142  & L-   &  R &	   &	      & 	     &  			    &	       &  LE    \\
132  & NGC4961      & 8185  & Sbc  &    & 2534(4)  & 93(3)    & 12.19(1.94)  & 4.98$\pm0.79\times 10^9$     &	       &  LE    \\
133  & A13065p5420  & 8231  & Sb   &  B &	   &	      & 	     &  			    &	       &  LE    \\
134  & IC4213	    & 8280  & Scd  &  B & 815(4)   & 81(4)    & 8.51(0.99)   & 4.44$\pm0.52\times 10^8$     &	       &  LE    \\
135  & A13194p4232  & 8400  & Scd  &  B &	   &	      & 	     &  			    &	       &  LE    \\
136  & NGC5117      & 8411  & Sc   &    & 2392(4)  & 96(3)    & 9.25(1.47)   & 3.51$\pm0.56\times 10^9$     &	       &  LE    \\
137  & NGC5173      & 8468  & E    &  B & 2428(6)  & 81(7)    & 4.99(1.06)   & 2.00$\pm0.42\times 10^9$     & $\times$ &  LE    \\
138  & A13281p3153  & 8498  & Sab  &  B & 7320(8)  & 290(19)  & 11.43(5.40)  & 3.19$\pm1.51\times 10^{10}$  &	       &  LE    \\
139  & NGC5208      & 8519  & L    &  R &	   &	      & $<$2.72	     & $<$6.36$    \times 10^9$	    &	       &  HI  \\
140  & NGC5230      & 8573  & Sc   &  B & 6856(6)  & 65(2)    & 8.75(1.39)   & 2.12$\pm0.34\times 10^{10}$  &	       &  LE    \\
141  & A13361p3323  & 8630  & Sm   &  B & 2438(7)  & 83(3)    & 6.70(0.58)   & 2.48$\pm0.22\times 10^9$     &	       &  LE    \\
142  & NGC5267      & 8655  & Sb   &  B &	   & 70(4)    & 0.78(0.16)   & 1.47$\pm0.30\times 10^9$     &	       &  LE    \\
143  & A13422p3526  & 8693  & Sbc  &  B & 2438(8)  & 106(4)   & 8.36(3.95)   & 3.41$\pm1.61\times 10^9$     &	       &  LE    \\
144  & NGC5338      & 8800  & L    &  R & 801(9)   & 20(1)    & 0.77(0.12)   & 1.93$\pm0.3 \times 10^7$     & $\times$ &  GBT   \\
145  & NGC5356      & 8831  & Sb   &  B & 1372(5)  & 122(4)   & 4.39(0.84)   & 6.75$\pm1.29\times 10^8$     &	       &  LE    \\
146  & A13550p4613  & 8876  & S0/a &  R &	   &	      & $<$0.15      & $<$4.78$    \times 10^7$     & $\times$ &  GBT   \\
147  & NGC5407      & 8930  & L    &  R &	   &	      & 	     &  			    &	       &  LE    \\
148  & NGC5425      & 8933  & Sc   &  B & 2072(8)  & 103(5)   & 6.64(0.78)   & 2.01$\pm0.24\times 10^9$     &	       &  LE    \\
149  & A14016p3559  & 8984  & L    &  R &	   &	      & 	     &  			    &	       &  LE    \\
150  & NGC5470      & 9020  & Sb   &    & 1026(8)  & 109(4)   & 7.69(1.47)   & 6.75$\pm1.29\times 10^8$     &	       &  LE    \\
151  & NGC5491      & 9072A & Sc   &  B & 5888(7)  & 214(3)   & 6.17(1.05)   & 1.10$\pm0.19\times 10^{10}$  &	       &  LE    \\
152  & NGC5532      & 9137  & L    &  R &	   &	      & 	     &  			    &	       &  LE    \\
153  & NGC5541      & 9139  & Sc   &  B &	   &	      & 	     &  			    &	       &  LE    \\
154  & NGC5596      & 9208  & L    &  R & 3220(24) & 94(7)    & 0.52(0.14)   & 3.17$\pm0.85\times 10^8$     & $\times$ &  GBT  \\
155  & NGC5608      & 9219  & Sm   &  B & 664(6)   & 46(2)    & 9.33(1.09)   & 4.19$\pm0.49\times 10^8$     &	       &  LE    \\
156  & A14305p1149  & 9356  & Sc   &  B & 2226(6)  & 103(3)   & 15.92(2.53)  & 5.18$\pm0.82\times 10^9$     &	       &  LE    \\
157  & NGC5684      & 9402  & L    &  R &	   &	      & 	     &  			    &	       &  LE    \\
158  & NGC5762      & 9535  & Sa   &  B & 1792(8)  & 80(3)    & 7.48(1.77)   & 1.71$\pm0.40\times 10^9$     &	       &  LE    \\
159  & A14489p3547  & 9560  & Pec  &  B & 1218(4)  & 80(4)    & 4.47(0.95)   & 5.99$\pm1.27\times 10^8$     &	       &  LE    \\
160  & A14492p3545  & 9562  & L+   &  B & 1257(3)  & 76(3)    & 11.97(2.29)  & 1.79$\pm0.34\times 10^9$     & $\times$ &  LE    \\
161  & IC1066	    & 9573  & Sab  &  B & 1576(6)  & 90(3)    & 7.69(2.93)   & 1.43$\pm0.54\times 10^9$     &	       &  LE    \\
162  & A14594p4454  & 9660  & Sc   &  B & 608(9)   & 29(2)    & 6.00(1.69)   & 2.75$\pm0.78\times 10^8$     &	       &  LE    \\
163  & A15016p1037  &	    & cE   &    &	   &	      & 	     &  			    &	       &  LE    \\
164  & IC1100	    & 9729  & Scd  &  B &	   &	      & 	     &  			    &	       &  LE    \\
165  & NGC5874      & 9736  & Sbc  &  B & 3128(5)  & 139(5)   & 6.82(1.08)   & 4.12$\pm0.65\times 10^9$     &	       &  LE    \\
166  & NGC5875A     & 9741  & Sc   &  B &	   &	      & 	     &  			    &	       &  LE    \\
167  & NGC5888      & 9771  & Sb   &  B &	   &	      & 	     &  			    &	       &  LE    \\
168  & IC1124	    & 9869  & Sab  &  B & 5313(8)  & 160(11)  & 3.48(0.90)   & 5.19$\pm1.34\times 10^9$     &	       &  LE    \\
169  & NGC5940      & 9876  & Sab  &    & 10214(6) & 78(3)    & 1.58(0.25)   & 7.90$\pm1.25\times 10^9$     &	       &  LE    \\
170  & A15314p6744  & 9896  & Sc   &    & 6466(6)  & 115(4)   & 4.34(0.42)   & 9.72$\pm0.94\times 10^9$     &	       &  LE    \\
171  & NGC5993      & 10007 & Sb   &  B & 9562(9)  & 76(6)    & 5.83(1.65)   & 2.67$\pm0.75\times 10^{10}$  &	       &  LE    \\
172  & IC1141	    & 10051 & S0/a &  B & 4389(3)  & 93(7)    & 1.82(0.07)   & 1.98$\pm0.08\times 10^9$     & $\times$ &  GBT   \\
173  & IC1144	    & 10069 & S0/a &  B &          &          & $<$0.07      & $<$5.37$     \times 10^8$    &          &  GBT   \\
174  & NGC6007      & 10079 & Sbc  &  B & 10544(5) & 184(3)   & 8.99(2.65)   & 4.89$\pm1.44\times 10^{10}$  &	       &  LE    \\
175  & A15523p1645  & 10086 & Sc   &  B &	   &	      & $<$2.06	     & $<$6.53$    \times 10^8$     &	       &  LE    \\
176  & A15542p4800  & 10097 & L    &  R &	   &	      & 	     &  			    &	       &  HI  \\
177  & NGC6020      & 10100 & ?E   &  R &	   &	      & $<$2.10	     & $<$2.26$    \times 10^9$	    &	       &  HI  \\
178  & NGC6123      & 10333 & S0/a &  R & 3964(12) & 34(3)    & 3.01(0.85)   & 2.78$\pm0.79\times 10^9$     &$\times$   &  LE    \\
179  & NGC6131      & 10356 & Sc   &  B & 5061(12) & 93(7)    & 4.85(1.37)   & 6.86$\pm1.94\times 10^9$     &	       &  LE    \\
180  & NGC6185      & 10444 & Sa   &  B &	   &	      & 	     &  			    &	       &  LE    \\
181  & NGC7077      & 11755 & S0/a &  B & 1148(6)  & 47(2)    & 2.00(0.41)   & 1.69$\pm0.35\times 10^8$     & $\times$ &  LE    \\
182  & NGC7194      & 11888 & ?E   &  R & 8105(?)  & 106(?)   & 6.17(0.58)   & 2.02$\pm1.90\times 10^9$	    &	       &  HI  \\
183  & A22306p0750  & 12074 & Sc   &  B & 1989(7)  & 70(3)    & 1.90(0.32)   & 4.15$\pm0.70\times 10^8$     &	       &  LE    \\
184  & NGC7328      & 12118 & Sab  &  B & 2824(6)  & 143(4)   & 10.33(1.75)  & 4.33$\pm0.73\times 10^9$     &	       &  LE    \\
185  & NGC7360      & 12167 & E    &  B & 4685(7)  & 157(8)   & 3.96(1.87)   & 4.31$\pm2.04\times 10^9$     & $\times$ &  LE    \\
186  & A22426p0610  & 12178 & Sdm  &  B & 1931(4)  & 103(7)   & 22.18(1.92)  & 4.46$\pm0.39\times 10^9$     &	       &  LE    \\
187  & A22551p1931N & 12265N& L    &  B & 5717(2)  & 103(7)   & 6.64(0.13)   & 1.07$\pm0.02\times 10^{10}$  & $\times$ &   GBT  \\
188  & NGC7436      & 12269 & E    &  R &	   &	      & 	     &  			    &	       &  LE    \\
189  & NGC7460      & 12312 & Sb   &  B & 3190(6)  & 90(3)    & 1.16(0.46)   & 6.28$\pm2.49\times 10^8$     &	       &  LE    \\
190  & NGC7537      & 12442 & Sbc  &  B & 2678(3)  & 142(3)   & 24.55(2.62)  & 8.74$\pm0.93\times 10^9$     &	       &  LE    \\
191  & NGC7548      & 12455 & L    &  R & 7991(6)  & 83(4)    & 0.29(0.07)   & 9.02$\pm2.18\times 10^8$     &	       &  LE    \\
192  & A23176p1541  & 12519 & Sd   &  B & 4378(5)  & 145(5)   & 4.23(0.32)   & 4.07$\pm0.31\times 10^9$     &	       &  LE    \\
193  & NGC7620      & 12520 & Scd  &  B & 9583(6)  & 108(3)   & 6.70(1.06)   & 3.11$\pm0.49\times 10^{10}$  &	       &  LE    \\
194  & A23264p1703  & 12620 & L    &  R & 6849     &	      & $<$0.16      & $<$3.69$    \times 10^8$     &	       &   CEGG  \\
195  & IC1504	    & 12734 & Sb   &  B & 6274(6)  & 204(7)   & 7.28(0.63)   & 1.39$\pm0.12\times 10^{10}$  &	       &  LE    \\
196  & NGC7752      & 12779 & Sd   &  B & 5044(8)  & 170(10)  & 8.67(1.95)   & 1.06$\pm0.24\times 10^{10}$  &	       &  LE    \\
197  & A23514p2813  & 12835 & E    &  R &          &          & $<$0.20      & $<$4.79$    \times 10^8$     &	       &   GBT  \\
198  & A23542p1633  & 12856 & Im   &  B & 1776(5)  & 66(2)    & 11.53(1.11)  & 2.02$\pm0.19\times 10^9$     &	       &  LE    \\
199  & A04345m0225  &  3104 & L    &    & 9739(8)  & 87(6)    & 1.37(0.39)   & 4.99$\pm1.42\times 10^9$     &	       &  LE    \\
200  & NGC1517      &  2970 & Sc   &    & 3483(3)  & 79(3)    & 10.71(0.93)  & 6.04$\pm0.52\times 10^9$     &	       &  LE    \\
\enddata 
\tablecomments{~Col.~(1): NFGS ID number. ~Col.~(2): NGC number, IC
  number, or IAU anonymous notation. ~Col.~(3): UGC number. ~Col.~(4):
  Morphological type in the NFGS database. ~Col.~(5): Sequence
  association (red, blue, or mid) if $M_*$ and $U-R$ color are
  available, see Figure \ref{fig.nfgssample}. ~Col.~(6): Heliocentric
  optical velocity measured from HI in ${\rm km\, s^{-1}}$. ~Col.~(7):
  Observed maximum rotation speed in ${\rm km\, s^{-1}}$. ~Col.~(8):
  Velocity-integrated HI flux in ${\rm Jy\, km\, s^{-1}}$. ~Col.~(9):
  HI gas mass calculated from $f_{\rm HI}$ following
  \citet{haynes84}. ~Col.~(10): Reference for HI data. GBT: new GBT
  observations presented in this paper, LE: HyperLeda
  \citep{paturel03b}, HR: {\it A General Catalog of HI Observations of
    Galaxies}, \citep{huchtmeier89}, CEGG: Cornell EGG HI Digital
  Archive \citep{springob05}, HI: HIPASS \citep{hipass01}.  }

\end{deluxetable}
\clearpage

\clearpage
\LongTables

\begin{deluxetable}{lllrrccccccc}

\tabletypesize{\scriptsize}
\tablewidth{0pt}
\tablecaption{New HI Data\label{table.newhi}}
\tablehead{
\colhead{ID}               
& \colhead{Galaxy Name}      
& \colhead{UGC}      
& \colhead{$\alpha_{2000}$}  
& \colhead{$\delta_{2000}$}  
& \colhead{$t_{int}$}  
& \colhead{$V_{\odot}$($\sigma_{V_{\odot}}$)}              
& \colhead{$W_{20}$($\sigma_{W_{20}}$)}    
& \colhead{$W_{50}$($\sigma_{W_{50}}$)}
& \colhead{$f_{\rm HI}(\sigma_{f_{\rm HI}}$)}                 
& \colhead{$\sigma_{\rm chan}$} 
\\
\colhead{}                 
& \colhead{}                 
& \colhead{} 
& \colhead{(J2000)}          
& \colhead{(J2000)} 
& \colhead{(s)}
& \colhead{(km\,${\rm s^{-1}}$)}        
& \colhead{(km\,${\rm s^{-1}}$)}           
& \colhead{(km\,${\rm s^{-1}}$)}
& \colhead{(Jy\,km\,${\rm s^{-1}}$)}     
& \colhead{(mJy)} 
\\
\colhead{(1)}               
& \colhead{(2)}      
& \colhead{(3)}      
& \colhead{(4)}  
& \colhead{(5)}  
& \colhead{(6)}  
& \colhead{(7)}              
& \colhead{(8)}    
& \colhead{(9)}
& \colhead{(10)}    
& \colhead{(11)}              
}

\startdata
1   & A00113+3037  & 130    & 00:13:56.9 & +30:52:59  & 2970  &           &	    &	     & $<$0.12       & 0.93  \\
11  & IC1639       & 750    & 01:11:46.5 & -00:39:52  & 2700  &  5381(12) & 83(35)  & 54(23) & 0.19(0.05)    & 1.00  \\
    &              &        &            &            &       &  5523(12) & 253(36) & 80(24) & 0.70 0.08)    & 1.00  \\
14  & NGC516       & 946    & 01:24:08.1 & +09:33:06  & 4440  &           &	    &	     & $<$0.09       & 0.80  \\
22  & IC195        & 1555   & 02:03:44.6 & +14:42:33  & 1140  &  3648(2)  & 264(7)  & 229(4) & 4.28(0.12)    & 1.46  \\
    &              &        &            &            &       &  3631(3)  & 474(10) & 393(7) & 6.85(0.16)    & 1.46  \\
30  & NGC1029      & 2149   & 02:39:36.5 & +10:47:36  & 870   &  3620(28) & 340(83) & 267(55)& 7.41(0.14)    & 1.56  \\
    &              &        &            &            &       &  3609(30) & 365(91) & 277(60)& 8.32(0.15)    & 1.56  \\
33  & NGC1298      & 2683   & 03:20:13.1 & -02:06:51  & 840   &  6452(3)  & 91(10)  & 86(6)  & 0.48(0.09)    & 1.87  \\
35  & NGC1552      & 2683   & 04:20:17.7 & -00:41:34  & 3180  &  4784(8)  & 318(24) & 281(16)& 0.70(0.09)    & 0.98  \\
44  & NGC3011      & 5259   & 09:49:41.2 & +32:13:15  & 2100  &  1543(2)  & 179(7)  & 170(5) & 1.24(0.08)    & 1.25  \\
54  & NGC3179      & 5555   & 10:17:57.2 & +41:06:51  & 1740  &           &	    &	     & $<$0.19       & 1.31  \\
68  & A10465+0711  & 5923   & 10:49:07.6 & +06:55:02  & 870   &  713(4)   & 188(11) & 149(7) & 3.12(0.16)    & 2.31  \\
69  & A10504+0454  & 6003   & 10:53:03.8 & +04:37:54  & 1740  &  5819(5)  & 234(16) & 152(11)& 1.47(0.09)    & 1.17  \\
72  & NGC3499      & 6115   & 11:03:11.0 & +56:13:18  & 6990  &  1495(12) & 370(35) & 261(23)& 0.39(0.06)    & 0.58  \\
75  & NGC3522      & 6159   & 11:06:40.4 & +20:05:08  & 900   &  1221(8)  & 308(24) & 246(16)& 2.00(0.16)    & 1.82  \\
80  & NGC3605      & 6295   & 11:16:46.6 & +18:01:02  & 900   &           &	    &	     & $<$0.11       & 1.33  \\
87  & A11332+3536  & 6570   & 11:35:49.1 & +35:20:06  & 3600  &  1628(2)  & 155(7)  & 113(4) & 1.35(0.05)    & 0.82  \\
93  & A11392+1615  & 6655   & 11:41:50.6 & +15:58:25  & 3600  &  744(2)   & 86(5)   & 63(3)  & 0.92(0.04)    & 0.97  \\
96  & A11476+4220  & 6805   & 11:50:12.3 & +42:04:28  & 3600  &  1132(6)  & 137(17) & 114(11)& 0.39(0.05)    & 0.90  \\
105 & A12001+6439  & 7020A  & 12:02:37.6 & +64:22:35  & 2100  &  1515(1)  & 136(4)  & 91(3)  & 2.30(0.05)    & 0.91  \\
106 & NGC4117      & 7112   & 12:07:46.1 & +43:07:35  & 900   &  950(2)   & 197(6)  & 197(4) & 2.20(0.13)    & 1.80  \\
    &              &        &            &            &       &  839(2)   & 480(7)  & 450(5) & 5.10(0.2)     & 1.80  \\
117 & NGC4308\tablenotemark{a} & 7426   & 12:21:56.9 & +30:04:27  & 8070  &           &    	    &	     & $<$0.06       & 0.61  \\
    &              &        &            &            &       &  702(5)   & 69(15)  & 56(10) & 0.13(0.03)    & 0.61  \\
144 & NGC5338      & 8800   & 13:53:26.5 & +05:12:28  & 900   &  801(9)   & 110(27) & 48(18) & 0.77(0.12)    & 2.23  \\
146 & A13550+4613  & 8876   & 13:56:58.0 & +45:58:24  & 2100  &           &	    &	     & $<$0.15       & 1.24  \\
154 & NGC5596      & 9208   & 14:22:28.7 & +37:07:20  & 1920  &  3220(24) & 385(73) & 209(49)& 0.52(0.14)    & 1.40  \\
172 & IC1141       & 10051  & 15:49:46.9 & +12:23:57  & 4140  &  4389(3)  & 237(8)  & 207(5) & 1.82(0.07)    & 0.92  \\
173 & IC1144\tablenotemark{b} & 10069  & 15:51:21.7 & +43:25:03  & 5700  &           &         &        & $<$0.07       & 0.62  \\
    &              &        &            &            &       &  12206(6) & 215(19) & 175(13)& 0.47(0.05)    & 0.62  \\
187 & A22551+1931N\tablenotemark{c} & 12265N  & 22:57:36.0 & +19:47:26  & 1140  &           &         &        & 1.66(0.13)    & 1.44  \\
    &              &        &            &            &       &  5717(2)  & 315(7)  & 227(5) & 6.64(0.13)    & 1.44  \\
197 & A23514+2813  & 12835  & 23:53:56.7 & +28:29:34  & 4080  &           &         &        & $<$0.20       & 0.79  \\
    &              &        &            &            &       &  6728(5)  & 204(16) & 182(10)& 0.43(0.06)    & 0.79  \\

\enddata

\tablecomments{For galaxies with large companions in the GBT beam, we
  measure the HI flux twice. The first row contains values measured in
  the ionized-gas/stellar velocity range, or upper limits if the
  galaxy is undetected. The second row includes companion flux (if
  any) within $\pm$300 ${\rm km\,s^{-1}}$ of the target galaxy. Please
  see \S \ref{section.companions} for more details. ~Col.~(1): NFGS ID
  number. ~Col.~(2): NGC number, IC number, or IAU anonymous
  notation. ~Col.~(3): UGC number. ~Col.~(4): RA of GBT pointing
  center in $hh$:$mm$:$ss.s$ (J2000). ~Col.~(5): DEC of GBT pointing
  center in $dd$:$mm$:$ss$ (J2000). ~Col.~(6): Time on-source in
  seconds. ~Col.~(7): Heliocentric optical velocity measured from HI
  in ${\rm km\, s^{-1}}$. ~Col.~(8): Velocity width measured at the
  20\% level in ${\rm km\, s^{-1}}$. ~Col.~(9): Velocity width
  measured at the 50\% level in ${\rm km\, s^{-1}}$. ~Col.~(10):
  Velocity-integrated HI flux in ${\rm Jy\, km\, s^{-1}}$. ~Col.~(11):
  Channel-to-channel RMS of the HI spectrum in mJy.  }

\tablenotetext{a}{The companion flux noted for this galaxy may in fact
belong to NGC~4308, although the offset between the optical and
measured HI velocity suggests that the flux belongs to a nearby
companion.}

\tablenotetext{b}{The measured HI velocity suggests that the flux
  belongs to a nearby companion.}

\tablenotetext{c}{Preliminary VLA HI data (Wei et al., in prep.)
  suggest that UGC~12265N contains about 1/4 of total HI within the
  GBT beam, so we use this fraction of the total HI flux for our
  analysis.}

\end{deluxetable}
\clearpage

\begin{deluxetable}{lll} 
  \tablecaption{Values of $M_*$ and $M_{\rm HI+He}/M_*$ for the sub-$M_b$ sample. \label{table.masses}}
  \tablehead{
    \colhead{Galaxy Name}
    & \colhead{$M_*$ ($M_{\odot}$)}
    & \colhead{$M_{\rm HI+He}/M_*$}}

  \startdata
  IC 1639    & $3.9\times 10^{10}$   & 0.01   \\
  IC 195     & $3.1\times 10^9$      & 0.12   \\
  NGC 3011   & $2.3\times 10^9$      & 0.12   \\
  UGC 6003   & $1.2\times 10^{10}$    & 0.29   \\
  IC 692     & $7.2\times 10^8  $    & 0.53   \\
  UGC 6637   & $1.6\times 10^9  $    & 0.38   \\
  UGC 6655   & $9.0\times 10^7$      & 0.26   \\
  UGC 6805   & $7.9\times 10^8  $    & 0.07   \\
  NGC 5173   & $1.9\times 10^{10}$    & 0.15   \\
  UGC 9562   & $7.5\times 10^8  $    & 3.35   \\
  IC 1141    & $2.3\times 10^{10}$    & 0.12   \\
  NGC 7077   & $6.8\times 10^8  $    &  0.34  \\
  NGC 7360   & $3.3\times 10^{10}$    & 0.18   \\
  UGC 12265N & $1.2\times 10^{10}$    & 0.30   \\
  \hline
  UGC 6570   & $3.6\times 10^9 $     & 0.10   \\
  UGC 7020A  & $2.2\times 10^9 $     & 0.25   \\
  \hline
  NGC 516    & $1.1\times 10^{10}$    & 0.003   \\
  UGC 5923   & $1.3\times 10^8$      & 0.51   \\
  NGC 3499   & $8.6\times 10^9$      & 0.01   \\
  NGC 3522   & $4.7\times 10^9$      & 0.07   \\
  NGC 3605   & $1.5\times 10^9  $    & 0.002   \\
  NGC 4117   & $5.0\times 10^9  $    & 0.05   \\
  NGC 4308   & $5.4\times 10^8  $    & 0.003   \\
  NGC 5338   & $7.2\times 10^8  $    & 0.04   \\
  UGC 8876   & $1.5\times 10^{10}$    & 0.004   \\
  NGC 5596   & $2.5\times 10^{10}$    & 0.03   \\
  NGC 6123   & $3.8\times 10^{10}$    & 0.10
  \enddata

  \tablecomments{Values of $M_*$ and $M_{\rm HI+He}/M_*$ for blue-,
    mid-, and red-sequence E/S0s in the sub-$M_b$ sample.}

\end{deluxetable}

\begin{deluxetable}{lllll} 

  \tablecaption{Fits of ${M_{\rm HI+He}/M_*}$ vs. ${M_*}$ \label{table.gsfits}}
  \tablehead{
    \colhead{NFGS population}
    & \colhead{Fit Type}
    & \colhead{Slope}
    & \colhead{Intercept}
    & \colhead{Scatter}}

  \startdata
  {\bf Blue Sequence:} &          &                &      &      \\
                       & forward  & -0.45 $\pm$ 0.05 & 4.09 & 0.45 \\
                       & bisector & -0.72          & 6.70 &      \\
  {\bf Spiral/Irregular:}     &          &                &      &      \\
                       & forward  & -0.45 $\pm$ 0.05 & 4.04 & 0.44 \\
                       & bisector & -0.70          & 6.49 &      \\
  {\bf Red Sequence:}  &          &                &      &      \\
                       & forward  & -0.67 $\pm$ 0.12 & 5.23 & 0.58 \\
                       & bisector & -1.46          & 13.0 &      \\
  {\bf E/S0:}          &          &                &      &      \\
                       & forward  & -0.67 $\pm$ 0.12 & 5.17 & 0.59 \\
                       & bisector & -1.57          & 13.9 &      \\
  \enddata

  \tablecomments{Fits of ${{\rm log}(M_{\rm HI+He}/M_*) = m\,{\rm
        log}(M_*) + b}$ for all galaxies in the NFGS with HI data,
    grouped by either sequence or morphological type. The forward fits
    were done using Buckley-James survival method in the ASURV package
    \citep{asurv}. The bisector fit is the mathematical bisector of
    the forward and backward fits using the same survival method, and
    so lacks estimates of the uncertainty and scatter.}

\end{deluxetable}

\newpage
\begin{deluxetable}{lllll} 

  \tablecaption{Timescales for Sub-${M_b}$ E/S0s\label{table.gastimes}}
  \tablehead{
    \colhead{Galaxy}
    & \colhead{Seq.}
    & \colhead{$\tau$}
    & \colhead{$t_{\rm dyn,HI}$}
    & \colhead{$t_{\rm dyn,GBT}$}
    \\
    \colhead{}
    & \colhead{}
    & \colhead{(Gyr)}
    & \colhead{(Gyr)}
    & \colhead{(Gyr)}}

  \startdata
  NGC 3011    & B   & 6.7   & 0.06  & 1.0    \\
  UGC 6003    & B   & 1.8   & 0.12  & 2.5    \\
  IC 692      & B   & 12.1  & 0.15  & 1.3    \\
  UGC 6637    & B   & 6.4   & 0.17  & 1.6    \\
  UGC 6655    & B   & 3.1   & 0.09  & 1.0    \\
  UGC 6805    & B   & 1.1   & 0.06  & 1.1    \\
  NGC 5173    & B   & 21.1  & 0.17  & 0.7    \\
  UGC 9562    & B   & 38.3  & 0.10  & 1.6    \\
  IC 1141     & B   & 2.5   & 0.07  & 1.3    \\
  NGC 7077    & B   & 2.7   & 0.17  & 1.1    \\
  NGC 7360    & B   & 168.5 & 0.15  & 1.1    \\
  UGC 12265N  & B   & 1.4   & 0.17  & 2.4    \\
  \hline
  UGC 6570    & Mid & 2.7   & 0.22  & 0.9    \\
  UGC 7020A   & Mid & 1.9   & 0.28  & 1.0    \\
  \hline
  UGC 5923    & R   & 18.7  & 0.04  & 0.7    \\
  NGC 5338    & R   & 0.8   & 0.40  & 0.4    
  \enddata
  
  \tablecomments{~Col.~(3): Gas exhaustion time, $\tau = M_{\rm
      HI+He}/{\rm SFR}$, in Gyr. ~Col.~(4): Inward travel time of gas
    from the edge of the HI disk, following $t_{\rm dyn} = \pi r_{\rm
      HI}/2v_{c}$ \citep{binney08}. We estimate $r_{\rm HI}$ using a
    typical ratio of HI to blue optical diameter of 2.11 for early
    type galaxies from \citet{noordermeer05}, and $v_c$ is the
    inclination-corrected maximum rotation speed of the HI from
    $V_{\rm M}^{{\rm sin} i}$. ~Col.~(5): Inward travel time of gas
    from the edge of the GBT beam, following $t_{\rm dyn} =
    \sqrt{\pi^2 r_{\rm beam}^3/4GM_{\rm tot}}$ \citep{binney08}, where
    $M_{\rm tot}$ is the stellar mass of the galaxy multiplied by 10
    to account for dark matter and $r_{\rm beam}$ is the distance from
    the edge of the GBT beam to the center of each galaxy.}

\end{deluxetable}

\clearpage
\begin{thebibliography}{87}
\expandafter\ifx\csname natexlab\endcsname\relax\def\natexlab#1{#1}\fi

\bibitem[{{Baldry} {et~al.}(2004){Baldry}, {Glazebrook}, {Brinkmann},
  {Ivezi{\'c}}, {Lupton}, {Nichol}, \& {Szalay}}]{baldry04}
{Baldry}, I.~K., {Glazebrook}, K., {Brinkmann}, J., {Ivezi{\'c}}, {\v Z}.,
  {Lupton}, R.~H., {Nichol}, R.~C., \& {Szalay}, A.~S. 2004, \apj, 600, 681

\bibitem[{{Barnes} {et~al.}(2001){Barnes}, {Staveley-Smith}, {de Blok},
  {Oosterloo}, {Stewart}, {Wright}, {Banks}, {Bhathal}, {Boyce}, {Calabretta},
  {Disney}, {Drinkwater}, {Ekers}, {Freeman}, {Gibson}, {Green}, {Haynes}, {te
  Lintel Hekkert}, {Henning}, {Jerjen}, {Juraszek}, {Kesteven}, {Kilborn},
  {Knezek}, {Koribalski}, {Kraan-Korteweg}, {Malin}, {Marquarding}, {Minchin},
  {Mould}, {Price}, {Putman}, {Ryder}, {Sadler}, {Schr{\"o}der}, {Stootman},
  {Webster}, {Wilson}, \& {Ye}}]{hipass01}
{Barnes}, D.~G., {et~al.} 2001, \mnras, 322, 486

\bibitem[{{Barnes}(2001)}]{barnes01}
{Barnes}, J.~E. 2001, in Astronomical Society of the Pacific Conference Series,
  Vol. 245, Astrophysical Ages and Times Scales, ed. T.~{von Hippel},
  C.~{Simpson}, \& N.~{Manset}, 382--+

\bibitem[{{Barnes}(2002)}]{barnes02}
{Barnes}, J.~E. 2002, \mnras, 333, 481

\bibitem[{{Bekki}(1998)}]{bekki98}
{Bekki}, K. 1998, \apjl, 502, L133+

\bibitem[{{Binney} \& {Merrifield}(1998)}]{binney98}
{Binney}, J., \& {Merrifield}, M. 1998, {Galactic Astronomy} (Galactic
  Astronomy : by James Binney and Michael Merrifield.~ Princeton, NJ :
  Princeton University Press, 1998.~ (Princeton series in astrophysics) QB857
  .B522 1998)

\bibitem[{{Binney} \& {Tremaine}(2008)}]{binney08}
{Binney}, J., \& {Tremaine}, S. 2008, {Galactic Dynamics: Second Edition}
  (Galactic Dynamics: Second Edition, by James Binney and Scott Tremaine.~ISBN
  978-0-691-13026-2 (HB).~Published by Princeton University Press, Princeton,
  NJ USA, 2008.)

\bibitem[{{Blitz} \& {Rosolowsky}(2004)}]{blitz04}
{Blitz}, L., \& {Rosolowsky}, E. 2004, \apjl, 612, L29

\bibitem[{{Bournaud} {et~al.}(2005){Bournaud}, {Jog}, \& {Combes}}]{bournaud05}
{Bournaud}, F., {Jog}, C.~J., \& {Combes}, F. 2005, \aap, 437, 69

\bibitem[{{Bower} {et~al.}(2006){Bower}, {Benson}, {Malbon}, {Helly}, {Frenk},
  {Baugh}, {Cole}, \& {Lacey}}]{bower06}
{Bower}, R.~G., {Benson}, A.~J., {Malbon}, R., {Helly}, J.~C., {Frenk}, C.~S.,
  {Baugh}, C.~M., {Cole}, S., \& {Lacey}, C.~G. 2006, \mnras, 370, 645

\bibitem[{{Ciotti} {et~al.}(1991){Ciotti}, {D'Ercole}, {Pellegrini}, \&
  {Renzini}}]{ciotti91}
{Ciotti}, L., {D'Ercole}, A., {Pellegrini}, S., \& {Renzini}, A. 1991, \apj,
  376, 380

\bibitem[{{Combes} {et~al.}(2007){Combes}, {Young}, \& {Bureau}}]{combes07}
{Combes}, F., {Young}, L.~M., \& {Bureau}, M. 2007, \mnras, 377, 1795

\bibitem[{{Cortese} \& {Hughes}(2009)}]{cortese09}
{Cortese}, L., \& {Hughes}, T.~M. 2009, \mnras, 1412

\bibitem[{{Crocker} {et~al.}(2008){Crocker}, {Bureau}, {Young}, \&
  {Combes}}]{crocker08}
{Crocker}, A.~F., {Bureau}, M., {Young}, L.~M., \& {Combes}, F. 2008, \mnras,
  386, 1811

\bibitem[{{Darg} {et~al.}(2009){Darg}, {Kaviraj}, {Lintott}, {Schawinski},
  {Sarzi}, {Bamford}, {Silk}, {Andreescu}, {Murray}, {Nichol}, {Raddick},
  {Slosar}, {Szalay}, {Thomas}, \& {Vandenberg}}]{darg09}
{Darg}, D.~W., {et~al.} 2009, ArXiv e-prints

\bibitem[{{Dasyra} {et~al.}(2006){Dasyra}, {Tacconi}, {Davies}, {Naab},
  {Genzel}, {Lutz}, {Sturm}, {Baker}, {Veilleux}, {Sanders}, \&
  {Burkert}}]{dasyra06}
{Dasyra}, K.~M., {et~al.} 2006, \apj, 651, 835

\bibitem[{{de Zeeuw} {et~al.}(2002){de Zeeuw}, {Bureau}, {Emsellem}, {Bacon},
  {Carollo}, {Copin}, {Davies}, {Kuntschner}, {Miller}, {Monnet}, {Peletier},
  \& {Verolme}}]{dezeeuw02}
{de Zeeuw}, P.~T., {et~al.} 2002, \mnras, 329, 513

\bibitem[{{Emsellem} {et~al.}(2004){Emsellem}, {Cappellari}, {Peletier},
  {McDermid}, {Bacon}, {Bureau}, {Copin}, {Davies}, {Krajnovi{\'c}},
  {Kuntschner}, {Miller}, \& {de Zeeuw}}]{emsellem04}
{Emsellem}, E., {et~al.} 2004, \mnras, 352, 721

\bibitem[{{Emsellem} {et~al.}(2007){Emsellem}, {Cappellari}, {Krajnovi{\'c}},
  {van de Ven}, {Bacon}, {Bureau}, {Davies}, {de Zeeuw}, {Falc{\'o}n-Barroso},
  {Kuntschner}, {McDermid}, {Peletier}, \& {Sarzi}}]{emsellem07}
---. 2007, \mnras, 379, 401

\bibitem[{{Faber} \& {Gallagher}(1976)}]{faber76}
{Faber}, S.~M., \& {Gallagher}, J.~S. 1976, \apj, 204, 365

\bibitem[{{Fakhouri} \& {Ma}(2008)}]{fakhouri08}
{Fakhouri}, O., \& {Ma}, C.-P. 2008, \mnras, 359

\bibitem[{{Fouque} {et~al.}(1990){Fouque}, {Durand}, {Bottinelli},
  {Gouguenheim}, \& {Paturel}}]{fouque90}
{Fouque}, P., {Durand}, N., {Bottinelli}, L., {Gouguenheim}, L., \& {Paturel},
  G. 1990, \aaps, 86, 473

\bibitem[{{Fraternali} \& {Binney}(2008)}]{fraternali08}
{Fraternali}, F., \& {Binney}, J.~J. 2008, \mnras, 386, 935

\bibitem[{{Governato} {et~al.}(2007){Governato}, {Willman}, {Mayer}, {Brooks},
  {Stinson}, {Valenzuela}, {Wadsley}, \& {Quinn}}]{governato07}
{Governato}, F., {Willman}, B., {Mayer}, L., {Brooks}, A., {Stinson}, G.,
  {Valenzuela}, O., {Wadsley}, J., \& {Quinn}, T. 2007, \mnras, 374, 1479

\bibitem[{{Hawarden} {et~al.}(1981){Hawarden}, {Longmore}, {Goss}, {Mebold}, \&
  {Tritton}}]{hawarden81}
{Hawarden}, T.~G., {Longmore}, A.~J., {Goss}, W.~M., {Mebold}, U., \&
  {Tritton}, S.~B. 1981, \mnras, 196, 175

\bibitem[{{Haynes} \& {Giovanelli}(1984)}]{haynes84}
{Haynes}, M.~P., \& {Giovanelli}, R. 1984, \aj, 89, 758

\bibitem[{{Hibbard} \& {van Gorkom}(1996)}]{hibbard96}
{Hibbard}, J.~E., \& {van Gorkom}, J.~H. 1996, \aj, 111, 655

\bibitem[{{Hopkins} {et~al.}(2009){Hopkins}, {Cox}, {Younger}, \&
  {Hernquist}}]{hopkins09}
{Hopkins}, P.~F., {Cox}, T.~J., {Younger}, J.~D., \& {Hernquist}, L. 2009,
  \apj, 691, 1168

\bibitem[{{Huchtmeier} \& {Richter}(1989)}]{huchtmeier89}
{Huchtmeier}, W.~K., \& {Richter}, O.-G. 1989, {A General Catalog of HI
  Observations of Galaxies. The Reference Catalog.} (A General Catalog of HI
  Observations of Galaxies.~The Reference Catalog.~Huchtmeier, W.K., Richter,
  O.-G., pp.~350.~ISBN 0-387-96997-7.~Springer-Verlag Berlin Heidelberg 1989)

\bibitem[{{Jansen} {et~al.}(2000{\natexlab{a}}){Jansen}, {Fabricant}, {Franx},
  \& {Caldwell}}]{jansen00a}
{Jansen}, R.~A., {Fabricant}, D., {Franx}, M., \& {Caldwell}, N.
  2000{\natexlab{a}}, \apjs, 126, 331

\bibitem[{{Jansen} {et~al.}(2000{\natexlab{b}}){Jansen}, {Franx}, {Fabricant},
  \& {Caldwell}}]{jansen00b}
{Jansen}, R.~A., {Franx}, M., {Fabricant}, D., \& {Caldwell}, N.
  2000{\natexlab{b}}, \apjs, 126, 271

\bibitem[{{Kannappan}(2004)}]{kannappan04}
{Kannappan}, S.~J. 2004, \apjl, 611, L89

\bibitem[{{Kannappan} \& {Fabricant}(2001)}]{kannappan01}
{Kannappan}, S.~J., \& {Fabricant}, D.~G. 2001, \aj, 121, 140

\bibitem[{{Kannappan} {et~al.}(2002){Kannappan}, {Fabricant}, \&
  {Franx}}]{kannappan02}
{Kannappan}, S.~J., {Fabricant}, D.~G., \& {Franx}, M. 2002, \aj, 123, 2358

\bibitem[{{Kannappan} \& {Gawiser}(2007)}]{kannappan07}
{Kannappan}, S.~J., \& {Gawiser}, E. 2007, \apjl, 657, L5

\bibitem[{{Kannappan} {et~al.}(2009){Kannappan}, {Guie}, \&
  {Baker}}]{kannappan09a}
{Kannappan}, S.~J., {Guie}, J.~M., \& {Baker}, A.~J. 2009, \aj, 138, 579 (KGB)

\bibitem[{{Kannappan} {et~al.}(2004){Kannappan}, {Jansen}, \&
  {Barton}}]{kannappan04a}
{Kannappan}, S.~J., {Jansen}, R.~A., \& {Barton}, E.~J. 2004, \aj, 127, 1371

\bibitem[{{Kannappan} \& {Wei}(2008)}]{kannappan08conf}
{Kannappan}, S.~J., \& {Wei}, L.~H. 2008, in American Institute of Physics
  Conference Series, Vol. 1035, The Evolution of Galaxies Through the Neutral
  Hydrogen Window, ed. R.~{Minchin} \& E.~{Momjian}, 163--168

\bibitem[{{Kannappan et al.}(2009)}]{kannappan09b}
{Kannappan et al.} 2009, in preparation, (K09b)

\bibitem[{{Kewley} {et~al.}(2006){Kewley}, {Geller}, \& {Barton}}]{kewley06a}
{Kewley}, L.~J., {Geller}, M.~J., \& {Barton}, E.~J. 2006, \aj, 131, 2004

\bibitem[{{Kewley} {et~al.}(2002){Kewley}, {Geller}, {Jansen}, \&
  {Dopita}}]{kewley02}
{Kewley}, L.~J., {Geller}, M.~J., {Jansen}, R.~A., \& {Dopita}, M.~A. 2002,
  \aj, 124, 3135

\bibitem[{{Knapp} {et~al.}(1978){Knapp}, {Kerr}, \& {Williams}}]{knapp78}
{Knapp}, G.~R., {Kerr}, F.~J., \& {Williams}, B.~A. 1978, \apj, 222, 800

\bibitem[{{Knapp} \& {Rupen}(1996)}]{knapp96}
{Knapp}, G.~R., \& {Rupen}, M.~P. 1996, \apj, 460, 271

\bibitem[{{Knapp} {et~al.}(1985){Knapp}, {Turner}, \& {Cunniffe}}]{knapp85}
{Knapp}, G.~R., {Turner}, E.~L., \& {Cunniffe}, P.~E. 1985, \aj, 90, 454

\bibitem[{{Krajnovi{\'c}} {et~al.}(2008){Krajnovi{\'c}}, {Bacon}, {Cappellari},
  {Davies}, {de Zeeuw}, {Emsellem}, {Falc{\'o}n-Barroso}, {Kuntschner},
  {McDermid}, {Peletier}, {Sarzi}, {van den Bosch}, \& {van de
  Ven}}]{krajnovic08}
{Krajnovi{\'c}}, D., {et~al.} 2008, \mnras, 390, 93

\bibitem[{{Lake} \& {Schommer}(1984)}]{lake84}
{Lake}, G., \& {Schommer}, R.~A. 1984, \apj, 280, 107

\bibitem[{{Lavalley} {et~al.}(1992){Lavalley}, {Isobe}, \& {Feigelson}}]{asurv}
{Lavalley}, M.~P., {Isobe}, T., \& {Feigelson}, E.~D. 1992, in Bulletin of the
  American Astronomical Society, Vol.~24, Bulletin of the American Astronomical
  Society, 839--840

\bibitem[{{Lees} {et~al.}(1991){Lees}, {Knapp}, {Rupen}, \&
  {Phillips}}]{lees91}
{Lees}, J.~F., {Knapp}, G.~R., {Rupen}, M.~P., \& {Phillips}, T.~G. 1991, \apj,
  379, 177

\bibitem[{{Leroy} {et~al.}(2008){Leroy}, {Walter}, {Brinks}, {Bigiel}, {de
  Blok}, {Madore}, \& {Thornley}}]{leroy08}
{Leroy}, A.~K., {Walter}, F., {Brinks}, E., {Bigiel}, F., {de Blok}, W.~J.~G.,
  {Madore}, B., \& {Thornley}, M.~D. 2008, \aj, 136, 2782

\bibitem[{{Li} {et~al.}(2008){Li}, {Kauffmann}, {Heckman}, {White}, \&
  {Jing}}]{li08}
{Li}, C., {Kauffmann}, G., {Heckman}, T.~M., {White}, S.~D.~M., \& {Jing},
  Y.~P. 2008, \mnras, 385, 1915

\bibitem[{{Li} {et~al.}(2006){Li}, {Mac Low}, \& {Klessen}}]{li06}
{Li}, Y., {Mac Low}, M.-M., \& {Klessen}, R.~S. 2006, \apj, 639, 879

\bibitem[{{Lin} \& {Tremaine}(1983)}]{lin83}
{Lin}, D.~N.~C., \& {Tremaine}, S. 1983, \apj, 264, 364

\bibitem[{{Marganian} {et~al.}(2006){Marganian}, {Garwood}, {Braatz},
  {Radziwill}, \& {Maddalena}}]{gbtidl}
{Marganian}, P., {Garwood}, R.~W., {Braatz}, J.~A., {Radziwill}, N.~M., \&
  {Maddalena}, R.~J. 2006, in Astronomical Society of the Pacific Conference
  Series, Vol. 351, Astronomical Data Analysis Software and Systems XV, ed.
  C.~{Gabriel}, C.~{Arviset}, D.~{Ponz}, \& S.~{Enrique}, 512--+

\bibitem[{{Morganti} {et~al.}(1997){Morganti}, {Sadler}, {Oosterloo},
  {Pizzella}, \& {Bertola}}]{morganti97}
{Morganti}, R., {Sadler}, E.~M., {Oosterloo}, T., {Pizzella}, A., \& {Bertola},
  F. 1997, \aj, 113, 937

\bibitem[{{Morganti} {et~al.}(2006){Morganti}, {de Zeeuw}, {Oosterloo},
  {McDermid}, {Krajnovi{\'c}}, {Cappellari}, {Kenn}, {Weijmans}, \&
  {Sarzi}}]{morganti06}
{Morganti}, R., {et~al.} 2006, \mnras, 371, 157

\bibitem[{{Naab} {et~al.}(2006){Naab}, {Jesseit}, \& {Burkert}}]{naab06}
{Naab}, T., {Jesseit}, R., \& {Burkert}, A. 2006, \mnras, 372, 839

\bibitem[{{Noordermeer} {et~al.}(2005){Noordermeer}, {van der Hulst},
  {Sancisi}, {Swaters}, \& {van Albada}}]{noordermeer05}
{Noordermeer}, E., {van der Hulst}, J.~M., {Sancisi}, R., {Swaters}, R.~A., \&
  {van Albada}, T.~S. 2005, \aap, 442, 137

\bibitem[{{Oosterloo} {et~al.}(2002){Oosterloo}, {Morganti}, {Sadler},
  {Vergani}, \& {Caldwell}}]{oosterloo02}
{Oosterloo}, T.~A., {Morganti}, R., {Sadler}, E.~M., {Vergani}, D., \&
  {Caldwell}, N. 2002, \aj, 123, 729

\bibitem[{{Paturel} {et~al.}(2003){Paturel}, {Theureau}, {Bottinelli},
  {Gouguenheim}, {Coudreau-Durand}, {Hallet}, \& {Petit}}]{paturel03b}
{Paturel}, G., {Theureau}, G., {Bottinelli}, L., {Gouguenheim}, L.,
  {Coudreau-Durand}, N., {Hallet}, N., \& {Petit}, C. 2003, \aap, 412, 57

\bibitem[{{Sadler} {et~al.}(2000){Sadler}, {Oosterloo}, {Morganti}, \&
  {Karakas}}]{sadler00}
{Sadler}, E.~M., {Oosterloo}, T.~A., {Morganti}, R., \& {Karakas}, A. 2000,
  \aj, 119, 1180

\bibitem[{{Sage} \& {Welch}(2006)}]{sage06}
{Sage}, L.~J., \& {Welch}, G.~A. 2006, \apj, 644, 850

\bibitem[{{Sage} {et~al.}(2007){Sage}, {Welch}, \& {Young}}]{sage07}
{Sage}, L.~J., {Welch}, G.~A., \& {Young}, L.~M. 2007, \apj, 657, 232

\bibitem[{{Sancisi}(1992)}]{sancisi92}
{Sancisi}, R. 1992, in Physics of Nearby Galaxies: Nature or Nurture?, ed.
  T.~X. {Thuan}, C.~{Balkowski}, \& J.~{Tran Thanh van}, 31--+

\bibitem[{{Schiminovich} {et~al.}(1995){Schiminovich}, {van Gorkom}, {van der
  Hulst}, \& {Malin}}]{schiminovich95}
{Schiminovich}, D., {van Gorkom}, J.~H., {van der Hulst}, J.~M., \& {Malin},
  D.~F. 1995, \apjl, 444, L77

\bibitem[{{Schneider} {et~al.}(1986){Schneider}, {Helou}, {Salpeter}, \&
  {Terzian}}]{schneider86}
{Schneider}, S.~E., {Helou}, G., {Salpeter}, E.~E., \& {Terzian}, Y. 1986, \aj,
  92, 742

\bibitem[{{Schneider} {et~al.}(1990){Schneider}, {Thuan}, {Magri}, \&
  {Wadiak}}]{schneider90}
{Schneider}, S.~E., {Thuan}, T.~X., {Magri}, C., \& {Wadiak}, J.~E. 1990,
  \apjs, 72, 245

\bibitem[{{Schweizer} \& {Seitzer}(1992)}]{schweizer92}
{Schweizer}, F., \& {Seitzer}, P. 1992, \aj, 104, 1039

\bibitem[{{Schweizer} {et~al.}(1989){Schweizer}, {van Gorkom}, \&
  {Seitzer}}]{schweizer89}
{Schweizer}, F., {van Gorkom}, J.~H., \& {Seitzer}, P. 1989, \apj, 338, 770

\bibitem[{{Serra} {et~al.}(2007){Serra}, {Trager}, {van der Hulst},
  {Oosterloo}, {Morganti}, {van Gorkom}, \& {Sadler}}]{serra07}
{Serra}, P., {Trager}, S.~C., {van der Hulst}, J.~M., {Oosterloo}, T.~A.,
  {Morganti}, R., {van Gorkom}, J.~H., \& {Sadler}, E.~M. 2007, New Astronomy
  Review, 51, 3

\bibitem[{{Skrutskie} {et~al.}(2006){Skrutskie}, {Cutri}, {Stiening},
  {Weinberg}, {Schneider}, {Carpenter}, {Beichman}, {Capps}, {Chester},
  {Elias}, {Huchra}, {Liebert}, {Lonsdale}, {Monet}, {Price}, {Seitzer},
  {Jarrett}, {Kirkpatrick}, {Gizis}, {Howard}, {Evans}, {Fowler}, {Fullmer},
  {Hurt}, {Light}, {Kopan}, {Marsh}, {McCallon}, {Tam}, {Van Dyk}, \&
  {Wheelock}}]{skrutskie06}
{Skrutskie}, M.~F., {et~al.} 2006, \aj, 131, 1163

\bibitem[{{Somerville} \& {Primack}(1999)}]{somerville99}
{Somerville}, R.~S., \& {Primack}, J.~R. 1999, \mnras, 310, 1087

\bibitem[{{Springob} {et~al.}(2005){Springob}, {Haynes}, {Giovanelli}, \&
  {Kent}}]{springob05}
{Springob}, C.~M., {Haynes}, M.~P., {Giovanelli}, R., \& {Kent}, B.~R. 2005,
  \apjs, 160, 149

\bibitem[{{Steinmetz} \& {Navarro}(2002)}]{steinmetz02}
{Steinmetz}, M., \& {Navarro}, J.~F. 2002, New Astronomy, 7, 155

\bibitem[{{Stewart} {et~al.}(2009){Stewart}, {Bullock}, {Wechsler}, \&
  {Maller}}]{stewart09}
{Stewart}, K.~R., {Bullock}, J.~S., {Wechsler}, R.~H., \& {Maller}, A.~H. 2009,
  ArXiv e-prints

\bibitem[{{Stewart} {et~al.}(2008){Stewart}, {Bullock}, {Wechsler}, {Maller},
  \& {Zentner}}]{stewart08}
{Stewart}, K.~R., {Bullock}, J.~S., {Wechsler}, R.~H., {Maller}, A.~H., \&
  {Zentner}, A.~R. 2008, \apj, 683, 597

\bibitem[{{Toomre} \& {Toomre}(1972)}]{toomre72}
{Toomre}, A., \& {Toomre}, J. 1972, \apj, 178, 623

\bibitem[{{van der Hulst} {et~al.}(2001){van der Hulst}, {van Albada}, \&
  {Sancisi}}]{vanderhulst01}
{van der Hulst}, J.~M., {van Albada}, T.~S., \& {Sancisi}, R. 2001, in
  Astronomical Society of the Pacific Conference Series, Vol. 240, Gas and
  Galaxy Evolution, ed. J.~E. {Hibbard}, M.~{Rupen}, \& J.~H. {van Gorkom},
  451--+

\bibitem[{{van Dokkum}(2005)}]{vandokkum05}
{van Dokkum}, P.~G. 2005, \aj, 130, 2647

\bibitem[{{van Gorkom} {et~al.}(1986){van Gorkom}, {Knapp}, {Raimond}, {Faber},
  \& {Gallagher}}]{vangorkom86}
{van Gorkom}, J.~H., {Knapp}, G.~R., {Raimond}, E., {Faber}, S.~M., \&
  {Gallagher}, J.~S. 1986, \aj, 91, 791

\bibitem[{{Verheijen} \& {Sancisi}(2001)}]{verheijen01}
{Verheijen}, M.~A.~W., \& {Sancisi}, R. 2001, \aap, 370, 765

\bibitem[{{Wardle} \& {Knapp}(1986)}]{wardle86}
{Wardle}, M., \& {Knapp}, G.~R. 1986, \aj, 91, 23

\bibitem[{{Welch} \& {Sage}(2003)}]{welch03}
{Welch}, G.~A., \& {Sage}, L.~J. 2003, \apj, 584, 260

\bibitem[{{White} \& {Frenk}(1991)}]{white91}
{White}, S.~D.~M., \& {Frenk}, C.~S. 1991, \apj, 379, 52

\bibitem[{{Young}(2002)}]{young02}
{Young}, L.~M. 2002, \aj, 124, 788

\bibitem[{{Young}(2005)}]{young05}
---. 2005, \apj, 634, 258

\bibitem[{{Young} {et~al.}(2008){Young}, {Bureau}, \& {Cappellari}}]{young08}
{Young}, L.~M., {Bureau}, M., \& {Cappellari}, M. 2008, \apj, 676, 317

\bibitem[{{Zaritsky} \& {Rix}(1997)}]{zaritsky97}
{Zaritsky}, D., \& {Rix}, H.-W. 1997, \apj, 477, 118

\end{thebibliography}
\end{document}